\documentclass[fleqn,usenatbib]{mnras}

\usepackage{txfonts}

\usepackage[T1]{fontenc}
\usepackage{supertabular}

\DeclareRobustCommand{\VAN}[3]{#2}
\let\VANthebibliography\thebibliography
\def\thebibliography{\DeclareRobustCommand{\VAN}[3]{##3}\VANthebibliography}

\usepackage{graphicx}	
\usepackage{pdflscape}
\usepackage{csvsimple}
\usepackage[export]{adjustbox}
\usepackage{makecell}
\usepackage{multirow}

\usepackage{wasysym} 

\title[{\textit{Gaia}}-\textit{GALEX} UV Excess Sample]{The white dwarf binary pathways survey - XI. The {\textit{Gaia}}-\textit{GALEX} UV Excess Sample for systems within 350\,pc}

\author[J. A. Garbutt et al.]{
James A. Garbutt,$^{1}$\thanks{E-mail: jagarbutt1@sheffield.ac.uk}
S. G. Parsons,$^{1}$
N. Gentile Fusillo$^{2}$
B. T. Gänsicke,$^{3}$
M. S. Hernandez,$^{4}$
\newauthor
F. Lagos-Vilches,$^{5}$
R. Raddi,$^{6}$
A. Rebassa-Mansergas,$^{6, 7}$
J. J. Ren,$^{8}$
M. R. Schreiber,$^{5}$
\newauthor
O. Toloza,$^{5}$
M. Zorotovic$^{9}$
\\
$^{1}$Astrophysics Research Cluster, School of Mathematical and Physical Sciences, University of Sheffield, Sheffield S3 7RH, UK\\
$^{2}$Department of Physics, Universita’ degli Studi di Trieste, Via A. Valerio 2, I-34127 Trieste, Italy\\
$^{3}$Department of Physics, University of Warwick, Coventry CV4 7AL, UK\\
$^{4}$Hamburger Sternwarte, University of Hamburg, Gojenbergsweg 112, 21029 Hamburg, Germany\\
$^{5}$Departamento de F{\'i}sica, Universidad T{\'e}cnica Federico Santa Mar{\'i}a, Avenida España 1680, Valpara{\'i}so, Chile\\
$^{6}$Departament de Física, Universitat Politècnica de Catalunya, c/Esteve Terrades 5, E-08860 Castelldefels, Spain\\
$^{7}$ Institut d'Estudis Espacials de Catalunya (IEEC), C/Esteve Terradas, 1, Edifici RDIT, 08860, Castelldefels, Spain\\
$^{8}$Key Laboratory of Space Astronomy and Technology, National Astronomical Observatories, Chinese Academy of Sciences, Beĳing 100101, P. R. China\\
$^{9}$Instituto de Física y Astronomía, Universidad de Valparaíso, Av. Gran Bretaña 1111, Valparaíso, Chile
}

\date{Accepted XXX. Received YYY; in original form ZZZ}

\pubyear{\the\year{}}

\begin{document}
\label{firstpage}
\pagerange{\pageref{firstpage}--\pageref{lastpage}}
\maketitle

\begin{abstract}
White dwarfs (WDs) with F, G or K type {binary companions} are the last common ancestor to a zoo of exotic phenomena. Finding a clean sample of such systems is difficult, as WDs in such binaries are usually outshone at optical wavelengths by their non-degenerate companion. It is useful to have a sample of such binaries - they can be studied to investigate binary evolution, as they may have undergone a previous phase of mass transfer. Using {\textit{Gaia}} data release 3 and \textit{GALEX}, we compile a sample of 2597 candidate WD\,+\,FGK systems within 350\,pc. These were identified via their ultraviolet excesses relative to MESA Isochrones and Stellar Tracks (MIST) model predictions consistent with the {\textit{Gaia}} parameters of their non-degenerate companions. We estimate the mass, surface gravity and effective temperature of the WDs by fitting the excess ultraviolet flux to models, or through interpolation where only one ultraviolet band was available. We also estimate the age of the system where possible, by comparing against any wide WD companions { in higher-order stellar systems}, or by checking if the system is a member of a cluster. {On} comparing against \textsc{Simbad}, we estimate the level of contamination within our sample {to be between 10-35 per cent.}. This is a valuable sample for investigating {WD\,+\,FGK binaries, particularly those containing} hotter WDs {($T_\mathrm{eff, WD} \gtrsim 25$\,kK)}, and {systems with post-main sequence companions}.
\end{abstract}




\section{Introduction} \label{sec:Intro}

White Dwarfs (hereafter WDs) are the final stage of evolution for a star of initial mass of $\lesssim$~9\,M$_\mathrm{\odot}$ \citep{Kepler_2007, Cummings_2018, Santos-Garcia_2025}. 18-26 per cent of WDs are known to exist in binaries \citep{Holberg_2009, Toonen_2017}, meaning that the WD previously evolved alongside a companion. If the initial orbital period of this binary system was short enough, $P_\mathrm{orb} \lesssim 10^4$\,d, then it is likely that the two stars will have interacted as the WD progenitor evolved \citep{Willems_2004}. Systems consisting of a WD and an M Dwarf (WD\,+\,dM) are well studied (e.g. \citealt{Rebassa-Mansergas_2021}) {-} those which are close enough will go on to become cataclysmic variables, otherwise they will remain detached. More interesting are systems consisting of a WD with a more massive F, G or K spectral type star (WD\,+\,FGK), as these have a more variable past and future evolution which we can probe.

We expect that systems containing an intermediate to high mass WD, $M_\mathrm{WD} \gtrsim 0.4$\,M$_\mathrm{\odot}$, with a shorter orbital period, $P_\mathrm{orb} \lesssim 100$\,d, to be the result of a phase of common envelope evolution \citep{Paczynski_1976, Zorotovic_2010}. Such a phase is difficult to model through hydrodynamics, though it generally involves a period of spiralling-in as friction robs the system of orbital energy. This phase generally occurs as the progenitor of the WD evolves onto the red giant branch (RGB) or onto the asymptotic giant branch (AGB), where the radius of the star dramatically increases causing the outer envelope to swallow the companion. Low mass, helium core WDs are thought to form through this pathway, as the outer layers of the RGB star are stripped away leaving behind the helium core as a WD, while higher mass CO or even ONe core WDs form during AGB interactions. An alternative method of interaction is stable mass transfer, wherein the companion to the WD progenitor accretes mass overflowing from the progenitor at a steady rate, which can lead to a widening of the binary \citep{Webbink_2008, Podsiadlowski_2014}. Until {\textit{Gaia}} data release 3 \citep{Gaia_2023}, the vast majority of WD\,+\,FGK binaries were short period systems of $\lesssim 2.5$\,d (e.g. \citealt{Parsons_2015, PathwayIV, PathwayVI, PathwayVII, PathwayVIII}). More recently, using {\textit{Gaia}}, longer period ($\gtrsim 100$\,d) WD\,+\,FGK binaries have been discovered (e.g. \citealt{Shahaf_2024, pathwayX}). 

There are three main methods for identifying WD\,+\,FGK binaries; astrometry (e.g. \citealt{Shahaf_2024, Yamaguchi_2025}), self-lensing (e.g. \citealt{Maeder_1973, Kruse_2014, Yamaguchi_2024}) and by their UV excess (e.g. \citealt{PathwayI, PathwayII, PathwayV}). In this paper, we will be using the latter method to create a sample of WD\,+\,FGK binaries using {\textit{Gaia}} data for the non-degenerate companion, and using \textit{GALEX} \citep{Martin_2005} UV measurements to detect potential hidden WDs. Previous samples using this method lacked {\textit{Gaia}} parallax measurements, and as such the parameters of the non-degenerate companion and of the binary system as a whole were thus less certain - and indeed subsequent observations of some of these systems (e.g. by \citealt{pathwayX}) were found to contain giant stars, when they were previously thought to be on the main sequence (MS). This may have caused some objects to be incorrectly flagged as having a UV excess, or genuine UV excess systems to be missed. A downside of the UV excess method is that it does not provide orbital separations, meaning detached WD\,+\,FGK binaries that will have never interacted may be present in the resulting sample. This method also is more sensitive to younger WDs, as these are hotter and thus more luminous at UV wavelengths.

We set out to make a new sample using both {\textit{Gaia}} data to identify the properties of the non-degenerate companion, and \textit{GALEX} data to probe for UV excesses, so that we can compile a sample with more certain constraints on the stellar companion (particularly effective temperature ($T_\mathrm{eff}$), radius ($R$), luminosity and metallicity), and thus a more accurate idea on if these systems have a genuine UV excess. We also determine correction factors to apply to MIST \footnote{\url{https://waps.cfa.harvard.edu/MIST/}} isochrones \citet{MIST0, MIST1, MESA_Stel_Astro, MESA_PORMS, MESA_BPE} model predictions in the UV, to bring them into agreement with observations and hence make it easier to identify genuine UV excess objects.

\section{Target Selection} \label{sec:TargetSelection}

The first step in defining our sample was to build a reliable {\textit{Gaia}}-\textit{GALEX} cross-match to serve as the basis for candidate selection. We started from the 1.4677 billion {\textit{Gaia}} DR3 sources with full astrometric solutions (i.e., measured parallax and proper motion) and cross-matched their J2016 coordinates with \textit{GALEX} GR7 \citep{Bianchi_2017}. For each {\textit{Gaia}} source, we retrieved all \textit{GALEX} detections within 40".

To account for proper motion, we compared the modified Julian date (MJD) of each \textit{GALEX} observation to {\textit{Gaia}}’s reference epoch (J2016; MJD 57 388). Using this epoch difference, we propagated the {\textit{Gaia}'s} positions backward to the \textit{GALEX} epoch for every candidate match. A {\textit{Gaia}}-\textit{GALEX} pair was considered a true match if the propagated {\textit{Gaia}} position lay within 3" of the \textit{GALEX} coordinates.

Because \textit{GALEX} has a significantly coarser spatial resolution (4.3\,-\,6.0" in $FUV$ and 5.6\,–\,8.0" in $NUV$) compared to {\textit{Gaia}} (0.4\,-\,0.5"), some \textit{GALEX} detections corresponded to multiple {\textit{Gaia}} sources. These cases likely reflect blended \textit{GALEX} sources, so we retained all of them. Conversely, some {\textit{Gaia}} sources matched multiple \textit{GALEX} detections, all of which corresponded to repeated \textit{GALEX} observations of the same object. Our final cross-matched catalogue contains 80.694 million {\textit{Gaia}}-\textit{GALEX} associations. {Having assembled a reliable catalogue of \textit{Gaia}-\textit{GALEX} crossmatches, we used it to create a sample of UV excess objects.}

\begin{figure*}
\centering
	\includegraphics[width = \linewidth]{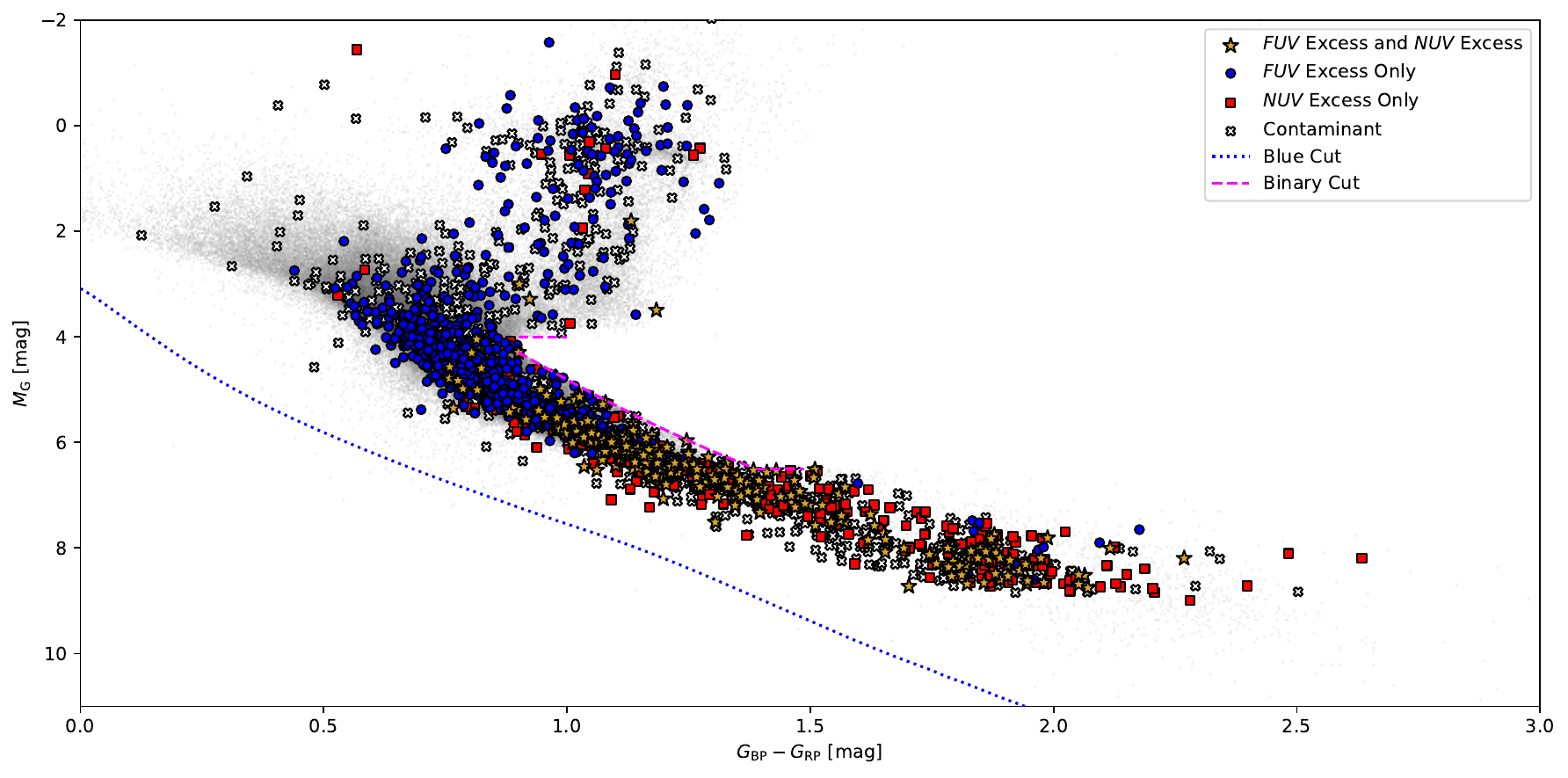}
	\caption{{\textit{Gaia}} optical colour magnitude diagrams displaying where our excess systems lie in  {\textit{Gaia}} optical colour space. Our systems with an excess in both UV bands are shown as golden stars, where there was only a $FUV$ excess as blue circles, systems with only an $NUV$ excess as red squares and likely contaminants as a white cross. These `Contaminants' are flagged based on the `Contaminants' column in the supplementary table. The blue dotted line represents to optical blue cut we made to remove lone WDs and most WD\,+\,dM binaries from our initial sample (criterion (5)), and the dashed magenta line represents the cuts we made to remove the binary track.}
	\label{fig:ExcessGaiaCMD}
\end{figure*}

In order to constrain this sample to both quality data and limit biases, we applied a number of {selection} criteria that each system had to meet. These are;
\begin{enumerate}
    \item[(1)] $\varpi$ > 0\,{mas},
    \item[(2)] $\varpi$/$\varpi$\_e > 10,
    \item[(3)] $NUV$\_e $\leq$ 0.1\,{mag},
    \item[(4)] {Sources have been detected in all three \textit{Gaia} bands},
	\item[(5)] {The} $G_\mathrm{BP} - G_\mathrm{RP}$ {colour} must be less than 0.5\,mag bluer than the solar-metallicity MIST zero-age main sequence {in the \textit{Gaia} optical CMD}(see dotted line in Figure~\ref{fig:ExcessGaiaCMD})
    \item[(6)] $d$ <= 350\,pc,
\end{enumerate}
where the distance $d$ is calculated by $1000/\varpi$. Criteria (1) and (2) ensure that we are working with systems {with} a robust parallax measurement, which {allows us to accurately} constrain {their} absolute magnitudes. Criterion (3) was applied so that we are only working with the systems with high precision $NUV$ data. We do not apply a similar cut in the $FUV$, as there are not as many systems with $FUV$ data owing to the $FUV$ instrument on \textit{GALEX} becoming inoperative {at the end of May 2009, four years before the end of the survey}. As a result of this instrumental failure, some of our systems lack $FUV$ data.

Criterion (4) was necessary for implementing criterion (5), which removed any obvious single WDs along with most WD\,+\,dM systems, which should lie bluer of this line as most WDs contribute significantly at optical wavelengths compared to a dM companion, and which are not what we are primarily looking for (for how to find such systems and their uses, see \citealt{Rebassa-Mansergas_2021}). A WD will contribute  negligible flux in optical wavelengths next to FGK stars, with a particularly hot WD ($\gtrsim 40$\,kK) next to a relatively cool K star being required to contribute 10 per cent of the $G_\mathrm{BP}$ flux.

{Criterion (6) sets the maximum distance at which a typical CO core WD of $T_\mathrm{eff}$ = 10000\,K and log($g$) = 8 will have a $NUV$ magnitude of 21\,mag; this criterion allows us to include also a significant number of giant stars in our sample.} We also removed {those matches with more than 4 \textit{GALEX} detections per \textit{Gaia} source, thus avoiding mismatches due to crowded fields.}

{We proceeded to apply the corrections of \citet{Wall_2019} to systems with an apparent magnitude brighter than 15.95\,mag in the $FUV$, and brighter than 16.95\,{mag} in the $NUV$. Subsequently, we accounted} for reddening using the extinction coefficient from {\textit{Gaia}} to de-redden our observed magnitudes. We used the \citet{Andrae_2023} {\textit{Gaia}} extinctions in order to remain self-consistent{, as we go on to use their stellar parameters.} {Given that we expect the optically luminous companion to dominate at optical wavelengths, these extinction estimates should not be significantly affected by any contributions from the WD, except for particularly hot WDs in excess of $\sim 40$\,kK next to a particularly cool K star}. We applied a further cut in order to remove the bulk of the binary track (this area will contain many active stars in binaries, a potential source of contamination). To make this cut, we removed systems that were redder than 0.5\,mag of the MS track as defined by equation~(6) of \citet{Kiman_2024}, bound by $4 \leq M_\mathrm{G} \leq 6.5${. This upper bound keeps evolved stars within our sample, whilst the lower bound keeps in K-type stars, as the MS track equation from \citet{Kiman_2024} does not perfectly fit our distribution dimmer than $\sim$\,6.5\,mag.} This cut, along with the cut of criterion (5) can be seen in Figure~\ref{fig:ExcessGaiaCMD}.

Finally, we apply {two} constraints to the non-degenerate star { based on the photometrically derived \textsc{gspphot} parameters of \citet{Andrae_2023}. First,} we remove systems where the log($g$) is greater than 4.6 in order to reduce the number of M dwarfs { in our excess sample. Second}, we remove systems where the effective temperature ($T_\mathrm{eff}$) is greater than 12,000\,K, in order to reduce {contamination from hotter,} more massive stars. After all of these cuts, we are left with a sample size of 488,148 systems, 111,292 of which have $FUV$ data.

One parameter we notably do not make cuts based around is $RUWE$ (Renormalized Unit Weight Error \citealt{Gaia_2023}), which is commonly used as an indicator of unresolved binarity, as it measures how well a given source matches a single star model. A non-optically luminous companion can cause an astrometric `wobble' in the photocentre of the binary, which would increase the measured $RUWE$ value, and which would affect parallax measurements. We however do not implement a cut on $RUWE$, as doing so would either remove very close systems (which have a low $RUWE$), or very wide systems (which have a high $RUWE$) - which would thus reduce the completeness of our sample, as we wish to include both in our sample.

\section{Method} \label{sec:Method}

\subsection{Defining the Excess Sample} \label{sec:DefExcess}


{In order to identify objects with UV excesses, we matched each of our systems against MIST isochrones containing \textit{Gaia} and \textit{GALEX} magnitudes. This match was performed using a Monte Carlo (MC) simulation using the \textit{Gaia} parameters for log($g$), metallicity (likewise \citealt{Andrae_2023}) and $T_\mathrm{eff}$, along with their errors. In each iteration, we matched the system to the isochrone with the closest metallicity (using isochrones with metallicities in the range of -4 to +0.5 in steps of 0.1) and interpolated the log($g$) and $T_\mathrm{eff}$ in order top acquire the intrinsic $FUV$ and $NUV$ magnitudes associated with these given parameters. Hereafter, `Observed' will be used to refer to the observed and dereddened absolute magnitude, and `Model' will refer to the intrinsic magnitudes that we interpolated from the MIST isochrones.}

\begin{figure*}
\centering
    \includegraphics[width =\linewidth]{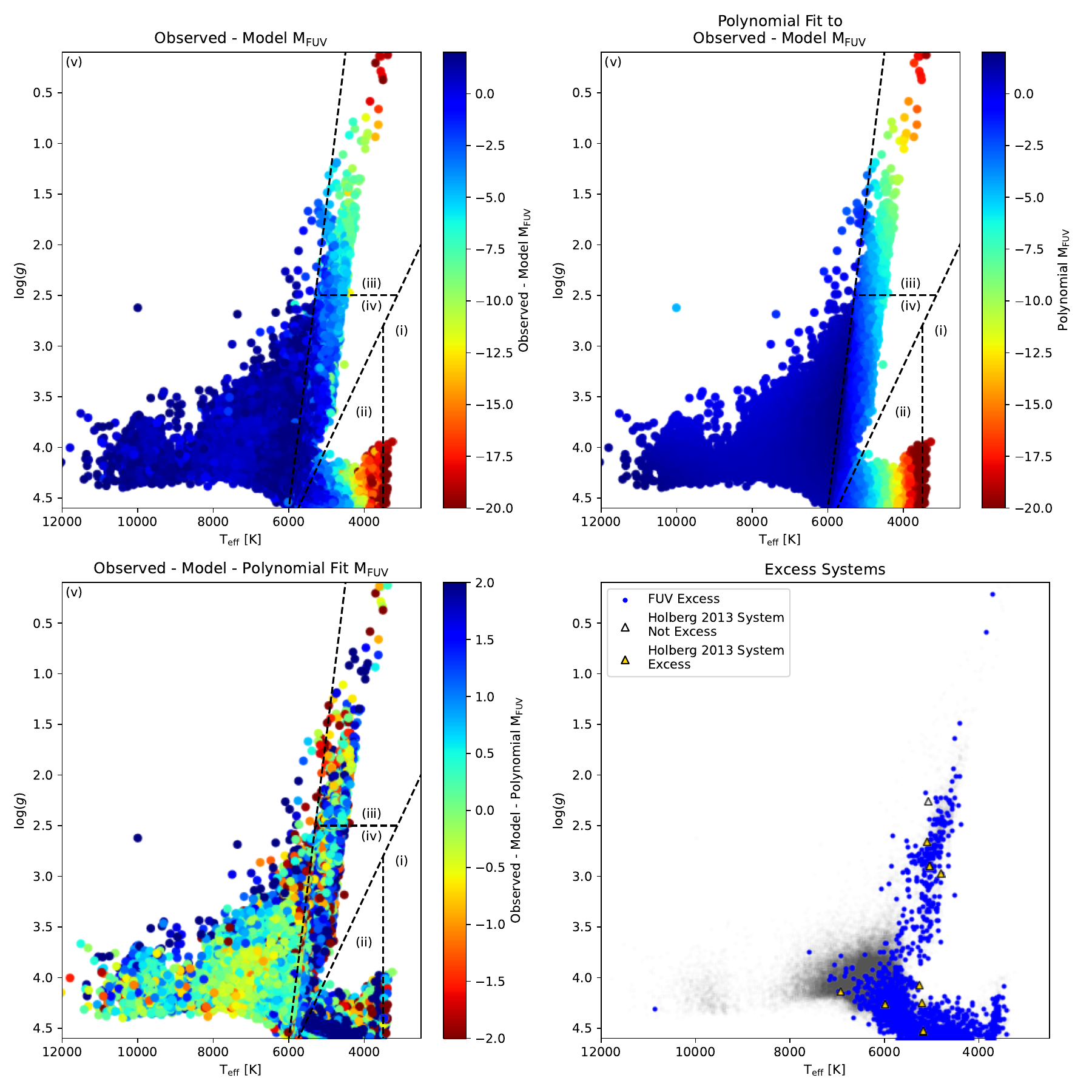}
    \caption{Kiel diagrams of the 111,292 objects in {\textit{Gaia}} DR3 with measured parameters, within 350\,pc and with a $FUV$ detection in \textit{GALEX}. The top left panel is colour-coded by the observed, absolute de-reddened $FUV$ magnitude minus the MIST model magnitude of the system as found by fitting the $T_\mathrm{eff}$, log($g$) and $\mathrm{Fe}/H$ of the system. The top right panel is colour-coded to show the polynomial fit to the distribution, and the bottom left shows the residuals after the polynomial corrections are subtracted. The dashed lines indicate the boundaries of the different regions where different polynomials were fit. The bottom right panel shows our systems that were flagged as an excess, with the known WD\,+\,FGK systems from \citet{Holberg_2013} shown as triangles.}
    \label{fig:loggTeffSpaceFUV}
\end{figure*}

A clear offset can be seen between the predicted UV magnitudes from the MIST models and the observed \textit{GALEX} UV magnitudes, as can be seen in the top-left panel in both Figure~\ref{fig:loggTeffSpaceFUV} and in Figure~\ref{fig:loggTeffSpaceNUV}. Whilst this could be an issue with the {\textit{Gaia}} parameters, this seems somewhat unlikely as this offset between observed and model UV fluxes has been {noted} in other models with non-{\textit{Gaia}} observed parameters (e.g. \citealt{PathwayVI, PathwayVII}). {It is more likely that this offset is a result of the chromosphere not being accounted for in the ATLAS12/SYNTHE synthetic photometry, used by the MIST models: chromospheric activity is most prominent at UV wavelengths in FGK stellar spectra \citep{Smith_2010}.} To correct for this, we fit a number of 2D polynomials to the distribution (top right panel) for both $FUV$ and $NUV$ bands assuming that the vast majority of the sample do not have a UV excess, but are representative of the typical UV fluxes of normal, isolated stars. These polynomials were fit in 5 different regions to best match the distribution, these regions being split by the following lines;

\begin{enumerate}
    \item[(a)] $T_\mathrm{eff}$ = 1250\,log($g$),
    \item[(b)] $T_\mathrm{eff}$ = (1000\,log($g$))\,/\,3\,+\,4466,
    \item[(c)] $T_\mathrm{eff}$ = 3500 when log($g$)\,$\geq$\,2.8,
    \item[(d)] log($g$) = 2.5 when 3125\,<\,$T_\mathrm{eff}$\,$\leq$\,5300,
\end{enumerate}

These zones thus correspond to;

\begin{enumerate}
    \item right of both lines (a) and (c),
    \item right of line (a) but left of line (c),
    \item left of line (a), right of line (b) and above of line (d),
    \item left of line (a), right of line (b) and below line (d),
    \item left of line (b).
\end{enumerate}

The resultant polynomial corrections to the {theoretical predictions}, given in the form
\begin{equation}
	A + B\,\mathrm{log(}g\mathrm{)} + C\,T_\mathrm{eff} + D\,\mathrm{log(}g\mathrm{)}^2 + E\,\mathrm{log(}g\mathrm{)}\,T_\mathrm{eff} + F\,T_\mathrm{eff}^2
\end{equation}
for each zone, with the coefficients $A - F$ given in Table~\ref{tab:PolyCoeff}.

\begin{table*}
	\centering
	\caption{The polynomial coefficients subtracted from the MIST models across the different zones in order to correct the model offsets. The polynomials take the form of $A + B\,\mathrm{log(}g\mathrm{)} + C\,T_\mathrm{eff} + D\,\mathrm{log(}g\mathrm{)}^2 + E\,\mathrm{log(}g\mathrm{)}\,T_\mathrm{eff} + F\,T_\mathrm{eff}^2$}
	\begin{tabular}{c|cccccc|cccccc}
		\hline
		\multirow{2}{*}{Zone} & \multicolumn{6}{c}{$NUV$ Polynomial Coefficients} & \multicolumn{6}{c}{$FUV$ Polynomial Coefficients}\\
		 & $A$ & $B$ & $C$ & $D$ & $E$ & $F$ & $A$ & $B$ & $C$ & $D$ & $E$ & $F$\\ 
		\hline
		(i) & 1665 & -248 & 6.68\,10$^{-1}$ & 6.26\,10$^{-1}$ & 7.19\,10$^{-2}$ & 5.42 10$^{-5}$ & 467 & -352 & 1.47\,10$^{-1}$ & 7.27 & 8.24\,10$^{-2}$ & -7.06\,10$^{-5}$\\
		(ii) & -12.5 & -28.5 & 2.74\,10$^{-2}$ & 2.42 & 1.91\,10$^{-3}$ & -3.48\,10$^{-6}$ & -324 & 97.9 & 3.38\,10$^{-2}$ & -12.4 & 2.54\,10$^{-3}$ & -3.81\,10$^{-6}$\\
		(iii) & 43.4 & 18.4 & -2.75\,10$^{-2}$ & 5.81\,10$^{-1}$ & -4.44\,10$^{-3}$ & 4.03\,10$^{-6}$ & -12.5 & 12.4 & -1.11\,10$^{-2}$ & 5.23\,10$^{-1}$ & 3.39\,10$^{-3}$ & 2.64\,10$^{-6}$\\
		(iv) & -4.13 & 4.02 & -8.93\,10$^{-4}$ & 1.47\,10$^{-1}$ & -9.52\,10$^{-4}$ & 4.15\,10$^{-7}$ & -126 & -6.02 & 4.56\,10$^{-2}$ & -1.63 & 3.16\,10$^{-3}$ & 4.77\,10$^{-6}$\\
		(v) & 11.3 & -3.79 & -4.17\,10$^{-4}$ & 4.57\,10$^{-1}$ & -6.85\,10$^{-5}$ & 4.22\,10$^{-8}$ & 8.28 & 5.03 & -4.65\,10$^{-3}$ & -1.18 & 6.76\,10$^{-4}$ & 1.08\,10$^{-7}$
	\end{tabular}
	\label{tab:PolyCoeff}
\end{table*}

When combined, these produce the residuals shown in the bottom-left panel of Figure~\ref{fig:loggTeffSpaceFUV} in the $FUV$ {band}. The { comparison plot for the} $NUV$ {band} (Figure~\ref{fig:loggTeffSpaceNUV}) looks similar, and in both cases no clear structure is seen in these residuals. We also investigated the effects of metallicity, though {we} found no systematic trends and thus we did not need to apply any corrections based on this factor. Finally, any systems that had an excess (Observed - Model - Polynomial) greater than a set value, $\Delta M_\mathrm{Limit}$ were flagged as being an excess in that band. {$\Delta M_\mathrm{Limit}$ was calculated by comparing each point in log($g$)\,/\,$T_\mathrm{eff}$ space with its 250 nearest neighbours which lie within 20 per cent of the parameter space. At each point, we calculate the median `Observed - Model - Polynomial' $FUV$ and $NUV$ magnitudes, and the 3 and 5 standard deviation values from the median for each band respectively. The mean of all of the 3 or 5 standard deviation values is taken as the $\Delta M_\mathrm{Limit}$ of the full sample in each band. $\Delta M_\mathrm{Limit, FUV}$ is found to be approximately 2.43\,mag, whilst $\Delta M_\mathrm{Limit, NUV}$ is found to be approximately 1.94\,mag.} We used the stricter 5\,$\sigma$ criterion for the $NUV$ band because some MS stars (especially F and early G) can contribute significantly in the $NUV$, so a stricter cut-off was necessary to account for this.

\begin{figure}
    \centering
    \includegraphics[width=0.9\linewidth]{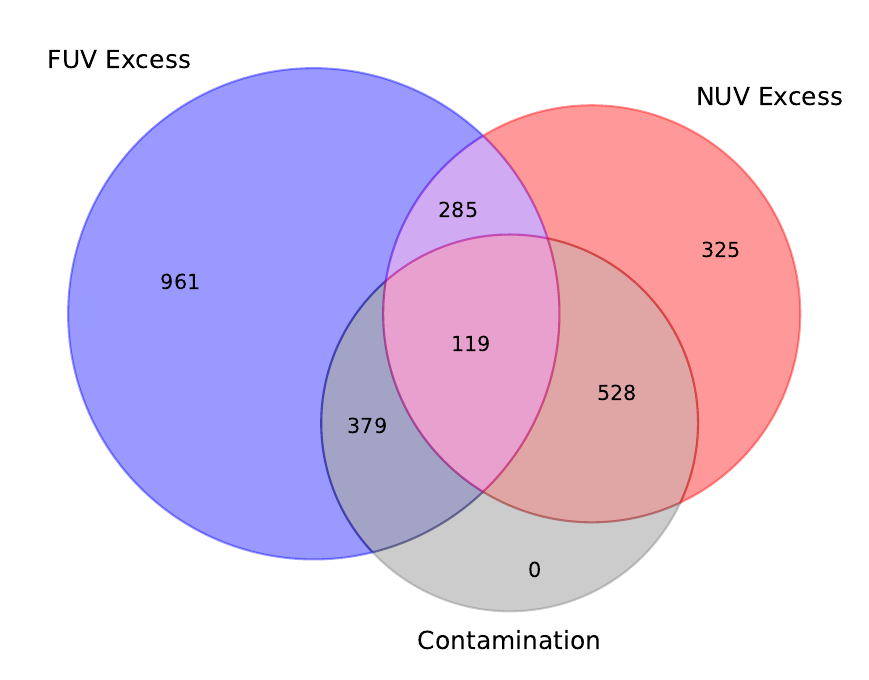}
    \caption{A set diagram showing the breakdown of which systems are flagged as having a $FUV$ and/or $NUV$ excess, along with the numbers of each which are flagged as contaminants (see Section~\ref{sec:Contamination}).}
    \label{fig:Excess_Set}
\end{figure}

We found a total of 1744 systems being flagged as having an $FUV$ excess, and 1257 having an $NUV$ excess, with 404 being common between the two as can be seen in Figure~\ref{fig:Excess_Set}, for a total excess sample size of 2597. Their locations in both {\textit{Gaia}} and UV Colour Magnitude Diagrams (CMDs) can be seen in Figures~\ref{fig:ExcessGaiaCMD} and \ref{fig:ExcessUVCMD}. {We find that 8 of the 9 of the previously known WD\,+\,FGK systems from \citet{Holberg_2013} that were in our sample after all the cuts and quality checks were flagged as an $FUV$ excess system, whilst of the 10 that were in the $NUV$ sample (as one did not have $FUV$ data), only 3 were flagged as having an $NUV$ excess. HD 13611 was not flagged in either band - though this is a relatively cool WD at $T_\mathrm{WD} \approx 13$\,kK \citep{Holberg_2013}, which would make it difficult to detect (see Section~\ref{sec:Completeness}).}

\subsection{Acquiring WD Parameters} \label{sec:WDParams}

With our sample of excess systems collated, we must now evaluate the parameters of our candidate WD - as it is possible that this UV excess may be from another source. Working under the initial assumption that the UV excess is caused by a WD, we {subtracted} the corrected {intrinsic} UV flux of the non-degenerate companion {(as found from the MIST isochrones and our derived correction factor} from the observed UV fluxes {in order} to obtain the {excess} UV flux. Overall, we have five scenarios;
\begin{enumerate}
    \item There is both an $FUV$ and $NUV$ excess in the system,
    \item There is a clear $FUV$ excess but only a marginal excess in the $NUV$,
    \item As above, but with the bands reversed,
    \item There is a clear $FUV$ excess, but no $NUV$ excess,
    \item There is a clear $NUV$ excess, but no $FUV$ excess, either because there is no $FUV$ data, or because there was no excess in the $FUV$.
\end{enumerate}
Here, `marginal' meaning that, whilst there is an excess in the band, it is too small to be attributed to a WD given the excess in the other band. In case (i), we perform a MC {simulation} using the UV magnitude{s} attributed to the WD and their respective uncertainties, fitting to both the DA CO core synthetic colour models of \citet{Bedard2020}\footnote{\url{https://www.astro.umontreal.ca/~bergeron/CoolingModels/}} and the He core models of \citet{Althaus_2013}\footnote{\url{https://evolgroup.fcaglp.unlp.edu.ar/modelos.html}} to extract log($g$), $T_\mathrm{eff}$ and $M_\mathrm{wd}$. If both models predict the mass to be below 0.5\,M$_\odot$, or if the CO core model could not return a value, we used the He core models. Otherwise, we used the CO core models. If {neither of the} models could return a fit, the method was flagged to have failed. The fit failing is {owed} to either the $NUV$ of $FUV$ magnitude falling outside the model grid, which could be due to an incomplete subtraction of the optically luminous companion, or the excess not being from a WD. {This method has resulted in a handful of systems where a fit could only be found in a few iterations, where the UV magnitudes and their uncertainties only overlap with a small region of the model grids. This has resulted in these systems having misleading errors on their output parameters (i.e. the values of log($g$), $T_\mathrm{eff}$ and $M_\mathrm{wd}$ for these systems appear more certain than is realistic). We use the DA models as we expect post-interaction systems to have a separation close enough that the WD will have accreted at least a small amount of hydrogen from their companion \citep{Parsons_2013}.} We fit here using log($g$) as a proxy for $R$ using the well tested mass-radius relationships of WDs (see e.g. \citealt{Holberg_2012, Tremblay_2017, Parsons_2017, Bedard_2017, Raddi_2025}).

In cases (ii) and (iii), we have fewer input parameters to work with.  We here instead run the MC {simulation} using the excess UV band and log($g$), allowing the fit to span the full range of log($g$) values for each model (7-9 for the CO core model, and $\sim$\,6\,-\,9.5 for the He core models). We discard the result of a MC iteration if the fit would have caused an excess in the band where we have only a marginal excess. This has a tendency to cause the fits to predict a higher mass, as a higher mass, fainter WD is easier to cause an excess in one band without contributing an excess in the other. As with case (i), we select which model is preferred based on the mass of the WD.

In cases (iv) and (v), we cannot do a proper fit. Instead, we interpolate using the band which has data and assuming a fixed log($g$) = 8 to find the $T_\mathrm{eff}$ and $M_\mathrm{WD}$. As with the other cases, we select which model is preferred based on the resultant mass of the fit. The spreads of the resultant parameters from cases (i) - (v) are given in Table~\ref{tab:MethodMeans}. Only the parameters of cases (i) - (iii) are given in the supplementary table.

The breakdown of each case can be observed in Table~\ref{tab:MethodBreakdown}. Not shown in the table are cases where the fitting methods failed. We find that 85 fail in case (i), 153 systems fail from case (ii), 70 fail in case (iii), 111 fail in case (iv), and 12 fail in case (v). These `failed' systems are owing to the parameters input in each case being {outside} the grid, and are flagged with flag `1' in the `Contamination' column of the supplementary table. These systems likely represent contaminants from active stars where the shape and strength of the UV excess is incompatible with a WD. However, this does not guarantee that all the rest of our sample is free from contamination as it is possible for activity to strongly mimic the SED of a WD at UV wavelengths (e.g. \citealt{pathwayX}).

\begin{table}
    \centering
    \caption{The number of systems fitted with each of our different fitting methods to determine the WD parameters.}
    \begin{tabular}{ll}
        \hline
        Fitting Method & Number \\
        \hline
        (i) Both Fit & 319\\
        (ii) Both Data, $FUV$ Excess Only & 756\\
        (iii) Both Data, $NUV$ Excess Only & 30\\
        (iv) $FUV$ Data Only & 320\\
        (v) $NUV$ Data Only & 741
    \end{tabular}
    \label{tab:MethodBreakdown}
\end{table}

\begin{figure*}
    \centering
    \includegraphics[height=0.9\textheight]{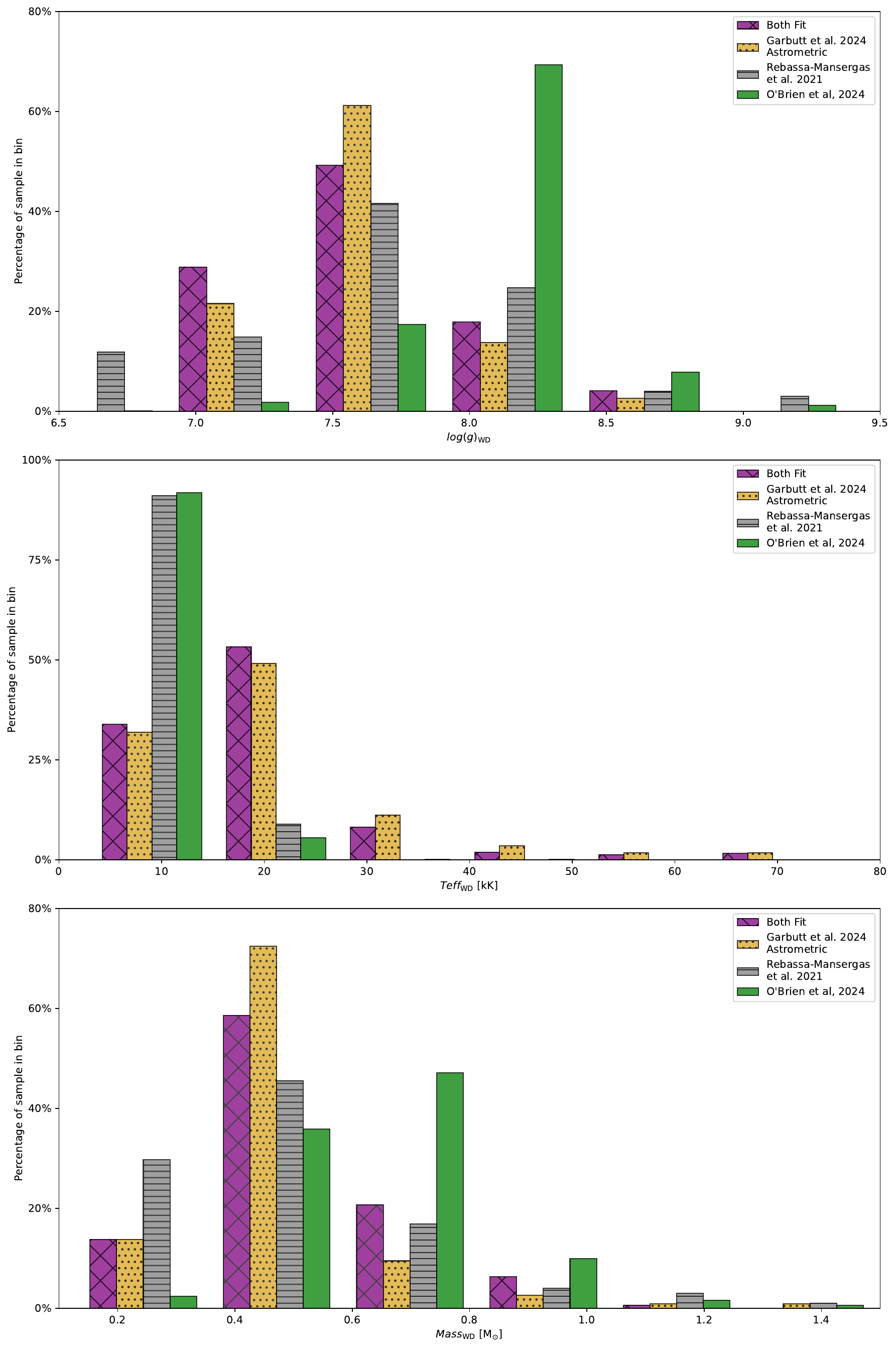}
    \caption{Histograms displaying the  case (i) WD parameters found from our fitting, shown in purple with crossed hatching, along with the same properties from the Astrometric sample of \citet{pathwayX} (as this sub-sample is more robust than the Spectroscopic sample of the same paper) shown in a yellow bar with a dotted hatch, the WD\,+\,dM sample of \citet{Rebassa-Mansergas_2021} shown as a grey bar with a horizontal hatch, and the 40\,pc sample of single WDs from \citet{O'Brien_2024}, shown as a solid green bar.}
    \label{fig:FitHist}
\end{figure*}

As can be seen in Figure~\ref{fig:FitHist}, when we consider our case (i) systems, we are most sensitive to lower mass WDs{. This} is unsurprising given that such WDs will be larger and more luminous, and thus more likely to have a UV excess. We compare our case (i) parameters against those from the astrometric sample of \citet{pathwayX}, the WD\,+\,dM sample of \citet{Rebassa-Mansergas_2021} and the 40\,pc single WD sample of \citet{O'Brien_2024}. We can see that, much like \citet{pathwayX} and \citet{Rebassa-Mansergas_2021}, we are sensitive to lower mass WDs, even more so than these samples. This result is expected in our sample, as we expect to find post-stable mass transfer binaries, which should have lower mass WDs - this is not the case in \citet{Rebassa-Mansergas_2021}, where their excess of very low mass systems is a result of their fitting method. This is in contrast to the 40\,pc single WD sample of \citet{O'Brien_2024}, which lacks the selection effects of the other samples in the comparison, which rely on more luminous WDs. However, there may still be a genuine difference in the mass distribution of binary vs single WDs, as binary interactions may result in mass loss.

\section{Crossmatching} \label{sec:Crossmatching}

With our excess samples acquired, we crossmatched with various external catalogues. The purposes of each crossmatch varied depending on the catalogue - from constraining the WD parameters better to identifying possible sources of contamination. A breakdown of the number of sources present in each crossmatch is given in Table~\ref{tab:CrossmatchNumbers}.

\begin{table}
    \centering
    \caption{A breakdown of the number of systems in each crossmatch. The radii given in the brackets next to \textsc{Simbad}, ROSAT and eROSITA are the sky-search radii used in crossmatching.}
    \begin{tabular}{ll}
        \hline
         Catalogue & Sample\,(2597)\\
         \hline
         Cluster Member & 13\,(1\%) \\
         Spatially Resolved Binaries & 741\,(29\%)\\
         \textsc{Simbad}\,(2") & 1618\,(62\%)\\
         {\textit{Gaia}} \verb|nss_acceleration_astro| & 161\,(6\%)\\
         {\textit{Gaia}} \verb|nss_non_linear_spectro| & 19\,(1\%)\\
         {\textit{Gaia}} \verb|nss_two_body_orbit| & 155\,(6\%)\\
         \textit{of which are astrometric} & 86\,(3\%)\\
         \textit{of which are single-lined spectra} & 62\,(2\%)\\
         ROSAT\,(10") & 126\,(5\%)\\
         eROSITA\,(2") & 66\,(3\%)\\
    \end{tabular}
    \label{tab:CrossmatchNumbers}
\end{table}

\subsection{Clusters}

It can be hard to constrain the age of WD\,+\,MS binaries post mass transfer as the two components have been affected by binary interactions, potentially erasing or resetting any age indicators. One way to constrain the ages of our systems would be if they belong to any clusters, as the population of the cluster should have the same age. To this end, we matched against the cluster catalogue of \citet{Hunt_2024}, finding that 30 of our systems are flagged as being a member of a cluster. A breakdown of these 13 systems can be seen in Table~\ref{tab:ClusterFlag} in Appendix~\ref{sec:SupTables}. Anything younger than 35\,Myr is too young to host a WD \citep{Yan_2025} and are flagged as `2' in the `Contamination' column in the supplementary table. For those clusters of around $\lesssim 10$\,Myr, these could potentially be young stellar objects \citep{Grankin_2016}.

\subsection{Spatially Resolved Binaries}

An alternate way of determining the age of the system is if there is a wide WD companion to the inner binary, i.e. a triple - as these stars would have all formed simultaneously, and finding the age of the distant WD that has not undergone binary interactions is fairly simple - simply adding the lifespan of the progenitor (based on the IMFM relationship) and the WD cooling age (see for example \citealt{Rebassa-Mansergas_2021}). We crossmatched with the catalogue of spatially resolved binaries of \citet{El-Badry_2021}, finding 741 matches. Not all of these are WD\,+\,MS binaries however - most are flagged as MS\,+\,MS binaries, meaning the wide companion to our inner binary is a MS star - for which it is harder to determine the age. 65 of these matches are flagged as WD\,+\,MS binaries.

It is possible that, of these 65, the UV flux that we have attributed to our inner binary has actually come from this wider WD. To check this, we matched the {\textit{Gaia}} source id of the wide companion back with our original {\textit{Gaia}}-\textit{GALEX} crossmatch table, and find 60 matches. All of the binaries with a angular separation, $\phi$, of $<\sim$5" have the same UV readings for both the inner object and the wide WD companion, suggesting that in these systems the entirety of the UV flux could come from this companion, and the inner object may indeed be a lone MS star. Of our whole excess sample, 652 of the 2597 (25 per cent) have another UV source within 5", with 631 of these having the same \textit{GALEX} object ID, though none of our excess objects are within 5" of one another. See Section~\ref{sec:Contamination} for a discussion of this potential contamination within our sample.

We have 7 systems with a WD common proper motion companion and $\phi$ greater than 5", though 3 of these lacked \textit{GALEX} data attributed to the wide companion's {\textit{Gaia}} source ID. Of the remaining 4 (WDJ082950.68+311257.30, WDJ083345.82+431745.64, WDJ085618.03+003758.87 and WDJ141407.54+831624.61), the difference between the UV magnitudes detected is promising that there is indeed an inner WD\,+\,MS binary in 2 of these systems, detailed in Table~\ref{tab:WideCompanion}.

\begin{table}
    \centering
    \caption{The 4 wide systems in our crossmatch with \citet{El-Badry_2021} with a distant WD companion to the UV excess object - too far away to be responsible for the UV excess, detailing their angular separation, the differences in the UV magnitudes between the wide companion and the inner binary, and the predicted age of the system.}
    \begin{tabular}{lllll}
        \hline
         WD & $\phi$\,["] & $\Delta M_\mathrm{FUV}$ & $\Delta M_\mathrm{NUV}$ & Age\,[Gyr]   \\
         \hline
         WDJ082950.68+311257.30 & 25.7 & 0.28 & 0.03 & -\\
         WDJ083345.82+431745.64 & 11.9 & - & 1.66 & 3.37\\
         WDJ085618.03+003758.87 & 5.3 & 0.47 & - & 4.17\\
         WDJ141407.54+831624.61 & 37.7 & 3.38 & 1.08 & -\\
    \end{tabular}
    \label{tab:WideCompanion}
\end{table}

We attempted to calculate the ages of these 4 systems using the {\sc{wdwarfdate}} python package \citep{Kiman_2022, Cummings_2018, Bedard2020, MIST0, MIST1}, assuming a DA WD and using the DA WD properties of \citet{WDsEDR3}. We find a cooling age of 3.73\,Gyr for WDJ083345.82+431745.64 and 4.17\,Gyr for WDJ085618.03+003758.87, though the package was unable to calculate an age for WDJ082950.68+311257.30 or WDJ141407.54+831624.61 due to the estimated very low masses of the WDs (each around 0.4\,M$_{\odot}$), outside the initial-final mass relation (IFMR) used by the model and thus it was unable to calculate the MS age for either system.

\subsection{{\textit{Gaia}} Non-Single Stars}

Given we are looking for what is optically a single source, but with a secondary hidden component, the {\textit{Gaia}} catalogues of such systems should provide us with a good match to what we are looking for, along with letting us see if we are looking at genuine close binaries opposed to chance alignments. To this end, we crossmatched our results with the {\textit{Gaia}} non-single stars catalogues (\verb|nss_acceleration_astro|, \verb|nss_non_linear_spectro|, \verb|nss_two_body_orbit| and \verb|nss_vim_fl|) to look for confirmed binaries. \verb|nss_acceleration_astro| is a catalogue where the optical source shows signs of astrometric acceleration that is best described by a quadratic or cubic fit, \verb|nss_non_linear_spectro| is similar, except for spectroscopic binaries opposed to astrometric binaries. \verb|nss_two_body_orbit| is the simplest, containing the linear fits of the binaries and \verb|nss_vim_fl| are systems where one of the binary components is a variable star, causing a shift in the photocentre of the system what required corrections to the astrometric parameters used in fitting the system \citep{Gaia_NSS}.

We find 161 matches with \verb|nss_acceleration_astro|, 19 in \verb|nss_non_linear_spectro|, 155 in \verb|nss_two_body_orbit|, and none in \verb|nss_vim_fl|. Of particular interest to us are the \verb|nss_two_body_orbit| results, as we can get a WD mass from these results as in \citet{pathwayX}. We find that of the 155 crossmatch results, 86 are astrometric systems and 62 are single-lined spectra, with the remaining 7 consisting of 6 double-lined spectroscopic binaries and one eclipsing binary, which are unlikely to be WD\,+\,FGK binaries, as a WD should contribute a negligible amount of optical flux compared to an F, G or K stellar type companion. A WD may contribute optically in a WD\,+\,dM binary, so in theory these 7 systems could have been a WD\,+\,dM, however inspecting the {\textit{Gaia}} colour magnitude diagram reveals that the eclipsing binary and 5 of the double lined spectroscopic binaries are F/G type MS stars, and the final double-lined spectroscopic system is an evolved star sitting above the horizontal branch - ruling out these systems being WD\,+\,dM systems.

\subsection{\textsc{Simbad}}

The biggest repository for us to check for contamination against would be \textsc{Simbad} \citep{SIMBAD}. We crossmatched with \textsc{Simbad} using a 2" sky search, converting our RA and dec to J2000 to do so. In particular, we were hoping \textsc{Simbad} would help us catch binary systems with an active star but no WD, pre-MS stars and poorly matched sources. In total, 1618 of our 2597 systems were successfully matched to a \textsc{Simbad} source.

To identify sources of contamination, we checked both the main type flagged by each object by \textsc{Simbad}, and the list of other types flagged. Broadly, we can split our types of interest into the following categories; Binary Stars, Erupting Variables, Galaxies, Planetary Nebulae, Planet Candidates, White Dwarfs and Young Stellar Objects. A summary of the breakdown of how many of each were flagged can be seen in Table~\ref{tab:Simbad_matches}.

\begin{table}
    \centering
    \caption{A breakdown of the broad categorizations of \textsc{Simbad} flags of interest. It should be noted that these categories can overlap, with systems being flagged as both `Binary' and as `Eclipsing Binary' separately. We have also not included in these numbers `candidate' systems, as these may not be contaminants.}
    \begin{tabular}{ll}
        \hline
        Flag & Sample\,(1618) \\
        \hline
        Binary Star & 173\,(11\%)\\
        Cataclysmic Variable & 1\,(0\%)\\
        Eclipsing Binary & 43\,(3\%)\\
        Erupting Variable & 9\,(1\%)\\
        Galaxies & 10\,(1\%)\\
        Planetary Nebulae & 2\,(0\%)\\
        Planet Candidate & 3\,(0\%)\\
        White Dwarfs & 16\,(1\%) \\
        Young Stellar Objects & 20\,(4\%)
    \end{tabular}
    \label{tab:Simbad_matches}
\end{table}

\subsubsection{Binary Stars}

\begin{figure*}
	\centering
	\includegraphics[width=\linewidth]{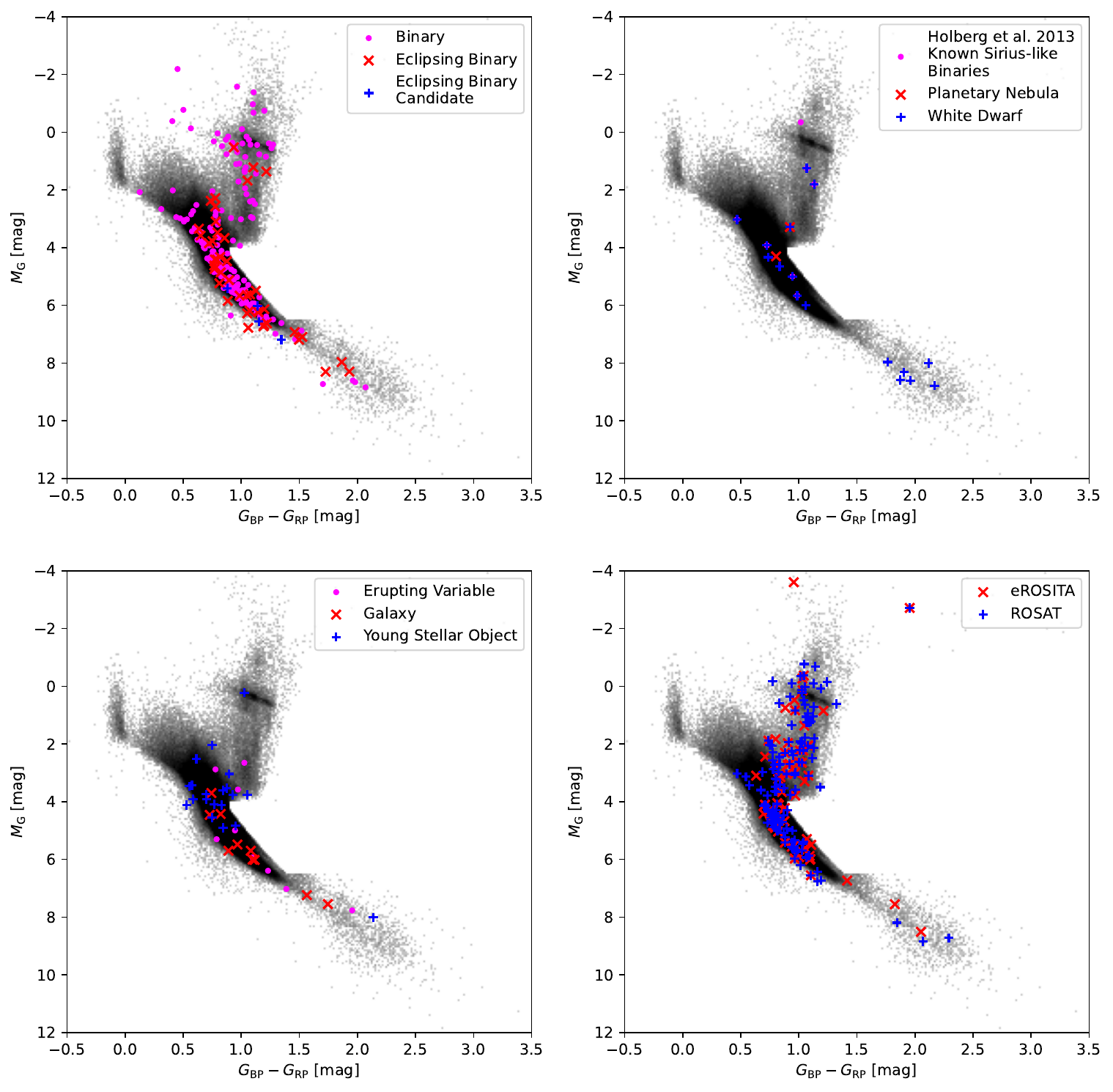}
	\caption{{\textit{Gaia}} optical colour-magnitude diagrams of our various crossmatches. The top left panel shows binaries (pink points), eclipsing binaries (red crosses) and eclipsing binary candidates (blue pluses) from our \textsc{Simbad} crossmatches, the top right panel shows the \citet{Holberg_2013} systems within our sample (pink points), along with the planetary nebulae (red crosses) and WDs found from our \textsc{Simbad} crossmatch. The bottom left panel shows the erupting variables (pink points), alleged galaxies (red crosses) and young stellar objects (blue pluses) from our \textsc{Simbad} crossmatch, and the bottom right shows the systems from our eROSITA crossmatch (red crosses) and ROSAT crossmatch (blue pluses).}
	\label{fig:SIMBAD_CMDs}
\end{figure*}

Given we are searching for WD\,+\,FGK binaries, one might think that things being flagged as a binary is a promising sign - however, it is not so simple, as we suspect that binary systems containing an active star may be a source of contamination, as chromospheric activity can cause an excess of UV flux \citep{Smith_2010, Shkolnik_2014, Smith_2018}. We made attempts to reduce the number of active MS\,+\,MS binaries by removing the binary track (see Section~\ref{sec:TargetSelection}), though it is likely that this remains a source of contamination within our final sample. We find that our flagged binaries cover the full range of the optical colour-magnitude diagram, from evolved stars to M dwarfs, as can be seen in Figure~\ref{fig:SIMBAD_CMDs}.

\subsubsection{Eclipsing Binaries}

Eclipsing binaries provide an indication that a system is unlikely to be a true WD\,+\,FGK candidate, as a WD does not contribute enough optical flux to undergo an obvious or deep eclipse, with the rare exception of an extremely hot WD next to a very late-type K dwarf, such as V* V471 Tau, a known WD\,+\,K binary which is found within our sample.

Indeed, we find 2 known WD binary candidates in our eclipsing sample that are not flagged as containing a WD by \textsc{Simbad}; CRTS J152748.4+353657 and V* V648 Aur. Eclipsing binaries (sans these 2 exceptions) are flagged as `3' in the `Contamination' column of the supplementary table.

CRTS J152748.4+353657 is disputed as being either a WD\,+\,dM system or being a neutron star (NS)\,+\,K system. This latter proposal is explored by \citet{Lin_2023}, whilst the former by \citet{Zhang_2024, Zhao_2024}. \citet{Zhang_2024} explains that this arises from \citet{Lin_2023} finding an inclination of $\sim\,45.2\deg$, whilst \citet{Zhang_2024} find an inclination of $\sim\,63\deg$, which led to the former getting a mass of 0.98\,M$_\mathrm{\odot}$, and the latter 0.69\,M$_\mathrm{\odot}$. \citet{Zhao_2024} independently found a mass of 0.67\,M$_\mathrm{\odot}$, in good agreement with \citet{Zhang_2024}. This mass combined with the UV excess leads us to conclude that the companion is most likely a WD.

V* V648 Aur is a RS Canum Venaticorum variable (RS CVn), a type of variable system defined by chromospheric activity within a close binary, with \citet{Lloyd_2007} identifying this system as being a short period ($\sim\,0.47$\,d) RS CVn with a K type star. It appeared as a candidate WD\,+\,FGK system in \citet{PathwayV}, however we cannot conclusively say whether this system contains a WD; only that it has been flagged as a candidate, since this type of active binary can be with or without a WD component.

\subsubsection{Galaxies}

We look here to figure out why we have seemingly flagged 10 galaxies - these could be genuine background galaxies, leading to UV contamination - or if the `Galaxy' classification is simply a miscategorisation. We flag these galaxies with a `4' in the `Contamination' column of the supplementary table. Images of these four galaxies can be seen in Figure~\ref{fig:Galactic_Images}.

2MASX J10500333+1538485 was matched with {\textit{Gaia}} DR3 3981520724472295168, which does not have its own entry within \textsc{Simbad} (2MASX J10500333+1538485 has its own DR3 identifier; {\textit{Gaia}} DR3 3981520720177408384. 2MASX J10500333+1538485 is flagged as a Seyfert 1 galaxy and as quasar \citep{Veron-Cetty_2010}. It is likely, given the differing DR3 identifiers and previous studies, that this is a background galaxy with {\textit{Gaia}} DR3 3981520724472295168 in the field.

2dFGRS TGS154Z168 was matched to {\textit{Gaia}} DR3 5120883704520804736, which itself does not have its own entry within \textsc{Simbad}. This was proposed in the 2dF Galaxy Redshift survey of \citet{Colless_2001}. Given that this survey was taken pre-{\textit{Gaia}}, and our reasonably robust parallax requirements, it is likely that this is a miscategorisation.

2dFGRS TGS392Z063 has a similar story to that of 2dFGRS TGS154Z168 - it was matched to {\textit{Gaia}} DR3 5067287254311288064, which again does not have it's own \textsc{Simbad} entry. Likewise a product of \citet{Colless_2001}, it it likely a miscategorisation.

ATO J214.6107+00.2412 was flagged as a galaxy by the Millennium Galaxy Catalogue \citep{Liske_2003}, however it is primarily flagged as a variable star, and is confirmed as a star by {\textit{Gaia}}. It is therefore another example of a previous miscategorisation.

LEDA 2094596 was matched with {\textit{Gaia}} DR3 760733820172658560 - which does not have its own entry in \textsc{Simbad}. LEDA 2094596 was initially flagged as a galaxy by \citet{HYPERLEDA}, and later by \citet{Moustakas_2023}. Given the recency of this latter study, we conclude that this is likely a background galaxy.

MCG-06-04-041 is a very obvious galaxy, as can be observed in panel 6 of Figure~\ref{fig:Galactic_Images} in Appendix~\ref{sec:SupImages}. This system has the correct {\textit{Gaia}} DR3 source ID - making it's inclusion interesting as a well studied galaxy (see e.g. \citealt{SouthernGalaxyCatalogue, Moustakas_2023}) with a robust {\textit{Gaia}} parallax placing it within 350\,pc. It is likely then that this is a background galaxy, where the star detected by {\textit{Gaia}} is positioned incredibly close to the galactic nuclei along {\textit{Gaia}}'s line of sight.

SDSS J093815.73+663622.8 appears in \citet{Toba_2014}, which flags it as a galaxy - however it is confirmed by {\textit{Gaia}} as being a star. We conclude then that this is yet another system previously miscategorised as a galaxy, given the robust parallax measurements of {\textit{Gaia}}.

TYC 4986-553-1 is primarily flagged as a star in \textsc{Simbad}. It appears in \citet{Liske_2003} and no other sources. As with our prior examples, we have a robust parallax for this system, so it is likely a miscategorisation.

UCAC4 234-000552 was investigated by the European Large Area ISO Survey \citep{Lari_2001, Rowan-Robinson_2004}, which is known to have both stellar and extragalactic sources. It has also been the target of the radial velocity experiment surveys \citep{RAVE4, RAVE5, RAVE6}, suggesting that this system is a star, and that the galaxy flag is a result of a miscategorisation.

Finally, [HM2015b] 96 1 was matched to {\textit{Gaia}} DR3 3981915444851588992, which does not have its own \textsc{Simbad} entry. [HM2015b] 96 1 is allegedly a UV source galaxy which is part of a star-forming compact group of galaxies \citep{Hernandez-Fern_2015}, though the parallax of this systems is robust, and appears in the {\textit{Gaia}} \verb|nss_acceleration_astro| catalogue. It being flagged as a galaxy is thus a miscategorisation of a star from an earlier misunderstanding of what the system was.

\subsubsection{White Dwarfs}

We find 16 WDs in our \textsc{Simbad} crossmatch. Most (10) of these are known WD binaries, or established WD binary candidates, detailed in Table~\ref{tab:SimbadWDs}. 6 systems however are currently flagged as lone WDs in \textsc{Simbad} - however these lie on the MS track in the {\textit{Gaia}} CMD (see the top-right panel of Figure~\ref{fig:SIMBAD_CMDs}). These are cases where \textsc{Simbad} has matched with a nearby resolved WD, which can be found in \citet{DR2WDs}. Five of these systems show up within our 5" search for nearby objects (see Section~\ref{sec:Contamination}, with the sixth being a proper common motion WD\,+\,MS binary, where the WD is around 6" away from the MS target). Two of our flagged WDs, BD+08 102 and PN A66 35, are discussed in more detail in the `Systems of Interest' section.

\subsubsection{Young Stellar Objects}

T Tauri variables produce a UV excess compared to MS stars \citep{Joy_1945, Joy_1954, Herbig_1962, 1970Esal, Sullivan_2022} as disc material accretes onto young star, and thus could be considered a potential source of contamination. However, these usually lie to the red of the MS track on the optical CMD, whilst those flagged in our sample are found spread over the full range of the optical CMD, as can be observed in Figure~\ref{fig:SIMBAD_CMDs}, with a slight over density sitting just below the MS, which we would not expect. We flag young stellar objects with the number `5' in the `Contamination' column of the supplementary table.

\subsubsection{Systems of Interest}

One of our systems, BD+08 102, is flagged as cataclysmic variable. However, BD+08 102 is unlikely to be a cataclysmic variable; it was first identified as a candidate WD\,+\,K binary by \citet{Kellett_1995}, and subsequent papers (such as \citealt{Leiner_2018, Sun_2024}), along with its appearance in {\textit{Gaia}}'s \verb|nss_acceleration_astro| catalogue suggest that it is a relatively wide binary. It is likely that this source was miscategorised as a cataclysmic variable owing to the presence of X-ray sources, which we instead believe to be coming from the rapidly rotating star, which is likely active.

We also found two apparent planetary nebulae - PN A66 35 and PN K 1-6. Such systems will have a UV excess caused by the now exposed core of the progenitor star of the nebulae, with the heated ejected material still providing more optical light than a lone WD. 

However, it is likely that PN A66 35 has been miscategorized and is not actually a planetary nebula, but rather an unrelated hot HII region ionized by a hot pre-WD (\citealt{Frew_2008, Tyndall_2013, Jacoby_2021}, under the name Abell 35 or A35), and can be seen in Figure~\ref{fig:PlanetaryNebulae_Images}. This pre-WD has only recently formed, and it is likely that the two progenitor stars had similar initial masses, given the companion star seems to be a G8 giant or subgiant nearing the end of its life, implying similar age \citep{Ziegler_2012}. Given the relatively long orbital period of $\sim$\,790\,d and their similar initial masses, this is a promising post-stable mass transfer candidate, as such systems are expected to form when the initial mass ratio is near 1, and are expected to emerge from the period of mass transfer with a relatively long orbital period. PN K 1-6 is itself a candidate for being a WD\,+\,G/K binary, though this is not the only proposed classification for this system - with it being a hot subdwarf binary also being likely \citep{Frew_2011}.

3 of our systems are flagged as potentially hosting exoplanets - HD 102888c, TOI-1893.01 and TOI-659.01, with these latter two being detected by their transits. We care about this as a WD transit and an exoplanet transit can mimic one another, making this a method for discovering WD binary candidates. In the case of TOI-659.01, the fit to the transit implies an object of $\sim 1$\,R$_\mathrm{Jup}$ - far too large to be a WD. TOI-1893.01 has a radius of $\sim 4$\,R$_\mathrm{\oplus}$, which could be consistent with a hot low mass WD - though more observations would be required to confirm the presence of a WD.

\subsection{ROSAT and eROSITA}

X-rays can be an indication for the presence of a WD in an interacting binary - as X-rays are produced during both accretion from Roche-lobe overflow, and also by wind accretion \citep{Kuijpers_1982, Patterson_1985, Mukai_2017}. However, they can also be produced by active stars - so the presence of X-rays is not confirmation that the system contains a WD, but is rather a flag for potential further follow up.

We matched with the second ROSAT all-sky survey \citep{ROSAT} and the SRG/eROSITA all-sky survey \citep{eROSITA} to see if we have matches with x-ray catalogues. We find 84 matched systems with the ROSAT survey using a 10" search radius, and 55 matches with the eROSITA survey using a 2" search radius. These matches seemingly span the full optical CMD, as seen in the bottom-right panel of Figure~\ref{fig:SIMBAD_CMDs}.

We find that our system flagged as a cataclysmic variable in \textsc{Simbad} does indeed appear in the ROSAT survey, though it does not appear in the eROSITA survey. We also find that 6 of the WDs flagged by \textsc{Simbad} appear in ROSAT (* 29 Dra, BD+08 102, CPD-65 264, HD 18131, HD 90052 and V* V471 Tau), and 2 appear in eROSITA (CPD-65 264 and TYC  110-755-1). Most are short period systems with active stars, so X-rays could originate from either or both components.

\subsection{Checks for Activity}

Another potential source of a UV excess within a system, aside from a hidden WD, is stellar activity. We make two checks for stellar activity. First, we checked against the catalogue of known active FGK stars of \citet{Gomes_da_Silva_2021}, though we find only one result - HD 16522. We also check for activity using \verb|activityindex_espcs| \citep{Creevey_2023}, a measure of stellar activity determined using Ca II emission. Following the method of \citet{Lanzafame_2023}, we find 143 of our systems are flagged as having very high activity, 544 as having a high activity, and 408 as being low activity. The remaining 1502 systems do not have a measurement of \verb|activityindex_espcs|, so we cannot measure activity using this method. We do not however flag the presence of activity within the `Contamination' column of the supplementary table, as the object could be a lone active star or an active star in a binary with a WD, as many active stars exist in binaries (see e.g. \citealt{PathwayVIII}).

\section{Completeness and Contamination} \label{sec:CompleteContamination}

\subsection{Completeness}
\label{sec:Completeness}

We do not expect that our sample of 2597 systems is complete. We are insensitive to cooler WDs, particularly next to hotter stars. To get an idea of what we are missing within our target range, we determined the maximum $T_\mathrm{eff, MS}$ for which a WD of given parameters can be detected, so we know which objects will have eluded our search. We are looking specifically at MS stars here, defined as having \verb|evolstage_flame| (a {\textit{Gaia}} parameter determining the life cycle phase of a stellar object using BaSTI models \citep{Hidalgo_2018, Creevey_2023}) $\leq$ 360, opposed to giants, as MS stars are more predictable than giant stars, and we do not have a statistically significant number of giant stars to compare against. 

In order to determine which WDs were detectable next to what MS star, we binned our MS stars that made it past our initial cuts in Section~\ref{sec:TargetSelection} prior to our UV excess selection by their $T_\mathrm{eff}$ in steps of 100\,K, but widening the bins in steps of 100\,K until these was at least 1000 systems in each bin. We averaged the $T_\mathrm{eff}$, log($g$), $M_\mathrm{NUV, Obs}$, $M_\mathrm{FUV, Obs}$, $M_\mathrm{NUV, MIST}$ and $M_\mathrm{FUV, MIST}$ in each bin, and used these to create a polynomial fit to the UV magnitudes.

\begin{figure*}
	\centering
	\includegraphics[width = \linewidth]{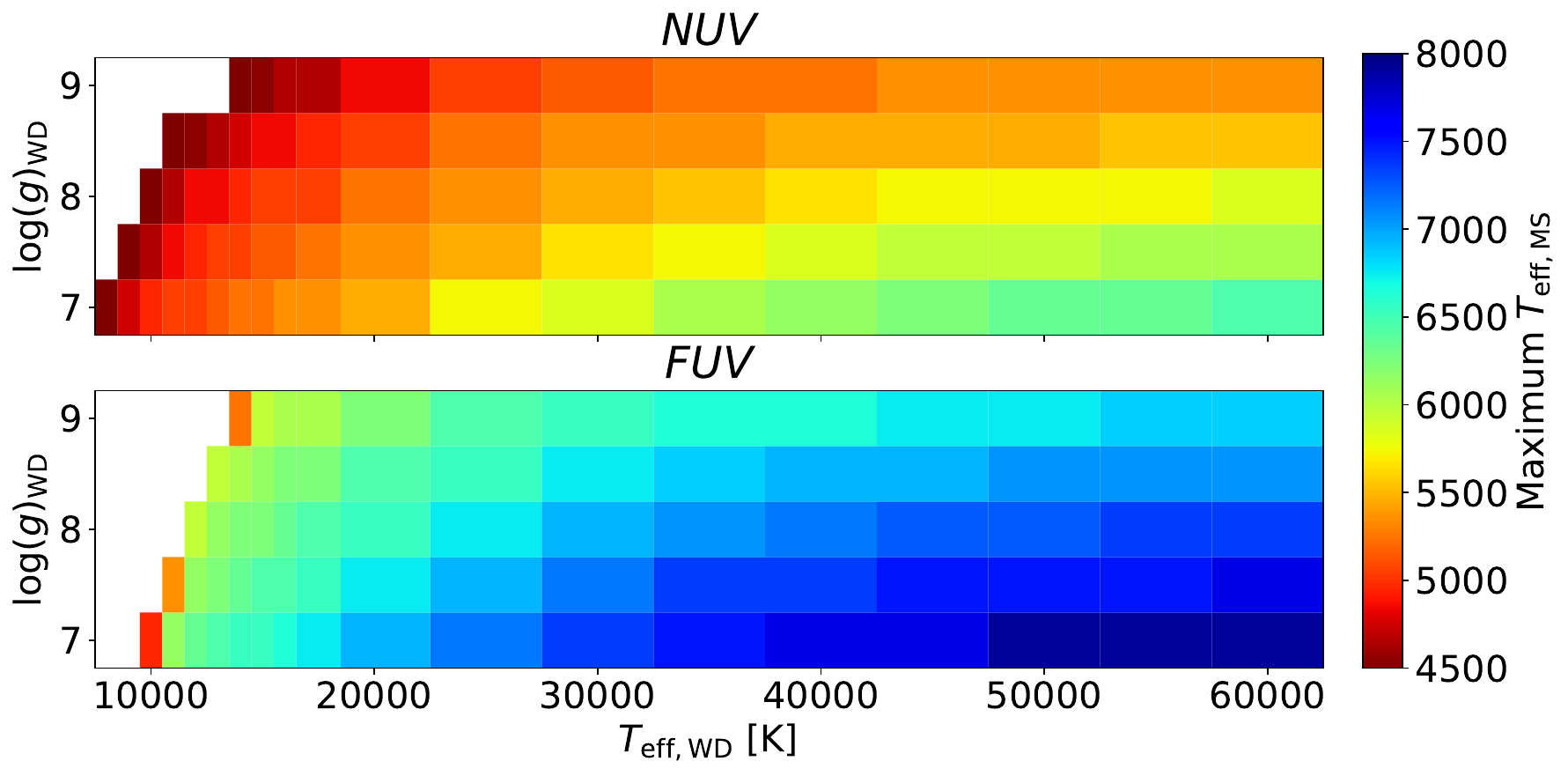}
	\caption{Grid showing the maximum $T_\mathrm{eff, MS}$ for which a WD of given $T_\mathrm{eff, WD}$ and log($g$) can be detected next to, for each the $NUV$ (top panel) and $FUV$ (bottom panel) bands. We find that we cannot detect particularly cool WDs next to even relatively cool MS stars, though as we look at hotted WDs, we can detect them next to even relatively hot MS stars in the $FUV$, and to a lesser extent in the $NUV$.}
	\label{fig:TeffLimit}
\end{figure*}

We compared a grid of CO core WD models of \citet{Bedard2020} against different temperatures of our MS stars, starting with the values of the hottest MS bin and progressively getting into cooler bins, until the WD became detectable as per our conditions in Section~\ref{sec:DefExcess}. As a result, we can determine the maximum $T_\mathrm{eff, MS}$ for which a WD with given parameters can be detected, as shown in Figure~\ref{fig:TeffLimit}. We expect a difference in the sensitivities between the $FUV$ and $NUV$ bands here, as WDs can produce more of an $FUV$ excess compared to even a hot MS star, which is not strictly the case regarding the $NUV$, as such we expect to be able to detect a WD in the $FUV$ next to hotter stars than is capable in the $NUV$.

As mentioned earlier in Section~\ref{sec:WDParams}, as with the previous sample of \citet{pathwayX}, we are more sensitive to lower mass ($M_\mathrm{WD} \lesssim 0.6$\,$\mathrm{M}_\mathrm{\odot}$), more luminous WDs. For a more complete catalogue, our sample could be combined with an astrometric counterpart (e.g. \citealt{Shahaf_2024}), which are more sensitive to higher mass WDs and biased against lower mass WDs. Yet to be explored are a population of low mass WDs next to more massive MS stars, as such systems would be difficult to detect though their UV flux - as such a MS star would also dominate in the UV regime as can be seen in Figure~\ref{fig:TeffLimit} where the higher $T_\mathrm{eff, MS}$ starts to dominate over the WD - and through astrometry, as it would be difficult to distinguish between a WD and a stellar companion next to such a luminous object.

We can check for further completeness by comparing against the TGAS (Tycho-{\textit{Gaia}} Astrometric Solution) survey of \citet{PathwayV}. The TGAS catalogue was constructed using {\textit{Gaia}} DR1 and the Tycho-2 catalogue, with \citet{PathwayV} complimenting this with the more up-to-date photometry and parallaxes of {\textit{Gaia}} DR2. Their sample of 814 DR2 sources becomes 846 DR3 sources owing to some DR2 sources being associated with multiple DR3 sources. Of these 846, 697 are found within our initial {\textit{Gaia}}-\textit{GALEX} sample. Applying our cuts from Section~\ref{sec:TargetSelection}, we find that none are dropped from criterion (1), 1 from criterion (2), 28 from criterion (3), 4 from criterion (4), 1 from criterion (5), and 199 from criterion (6), leaving us with a sample of 464 candidates from \citet{PathwayV} that have passed our initial filters, with a further 96 not having both log($g$) and $T_\mathrm{eff}$ within our range and 9 being removed by our binary cut, for a sample size of 359.

We find that 233 of these 359 (65 per cent) possible sources from \citet{PathwayV} are flagged as UV excess sources via our method, leaving 126 apparent excess sources from \citet{PathwayV} that pass all our quality cuts but are not flagged as having a UV excess. The majority of these objects are hot A/F-type MS stars (The average $T_\mathrm{eff}$ of these systems is 6600\,K), which we are not sensitive to a UV excess from, except in the case of a very hot WD companion (see Figure~\ref{fig:TeffLimit}). We suspect that these are likely not real UV excess sources, but just hot MS stars (that will be intrinsically bright at UV wavelengths) that were erroneously flagged by \citet{PathwayV}.

It would be near impossible to get a genuine measure of how complete our sample is without employing a population synthesis model, however population synthesis has thus far proven unreliable for post-interaction WD binaries, particularly regarding stable mass transfer and double WD binaries \citep{Lin_2023, Nelemans_2025}. \citet{PathwayI} shows in their population synthesis that we should not expect any cool WDs ($T_\mathrm{eff, WD} \lesssim$ 15000\,K) next to a hot MS star ($T_\mathrm{eff, MS} \gtrsim$ 6500\,K). This is to be expected, as such a hot MS star should have evolved and become a WD itself by the time the WD has formed and cooled to this temperature. However, it is apparent that our sample becomes significantly incomplete below $T_\mathrm{eff, WD} \lesssim 12,000$\,K. We are also limited by the lack of full-sky $FUV$ coverage, as the \textit{GALEX} $FUV$ instrument failed before it could complete full coverage.

Whilst there is no overlap, the WD\,+\,dM survey of \citet{Rebassa-Mansergas_2025} should prove to be a complimentary survey to ours - as they probe a different area of the {\textit{Gaia}} optical CMD (they search between the WD track and the MS track). These systems exist in a different area of parameter space, as a cooler WD can contribute significantly in optical wavelengths when in a binary with a dM star.

\subsection{Contamination}
\label{sec:Contamination}

By comparing against our \textsc{Simbad} crossmatch, we can get an approximate measure of contamination within our sample, measuring approximately 5 per cent depending on how representative the \textsc{Simbad} sub-sample is relative to the full sample. We get this number by considering 41 of the 43 eclipsing binaries as contaminants, all 10 of the `galaxies' as contaminants, and the 20 young stellar objects as contaminants.

We expect a few sources of contamination within our sample beyond these kinds of objects. We expect the primary source of contamination within our sample to be systems containing an active star, as chromospheric activity can produce UV flux that would register within our detection parameters. We made an effort to reduce this source of contamination by removing the binary track, as this is where we expect a large number of these systems to lie, but there is still likely more in the sample, and removing all active stars would be counter-productive, as WD\,+\,FGK binaries can have stellar activity.

We also expect contamination from genuine multiple-star systems involving a WD, such as triple systems where there is a wide WD companion (e.g. \citealt{PathwayIII}), or WD binaries that have not undergone interactions owing to having too wide of an orbit. Such systems will, of course, slip through our filters and enter our sample - as the presence of a WD with a non-degenerate companion is exactly what our filters select for, and will require follow-up data to confirm.

Additionally, there could be contamination from foreground or background UV sources. To check for this, we checked our excess sample against the full initial {\textit{Gaia}}-\textit{GALEX} crossmatch table, searching around each source within a 5" radius. We find that 652 of our 2597 excess systems have another source within 5", with 631 of these sharing a \textit{GALEX} object ID. We flag our sources with another detection within 5" where the other detection is unlikely to be the cause of our UV excess with a `6' in the `Contamination' column of the supplementary table.

Of these 652 systems, we find that 533 lie across the MS and giant branch of the {\textit{Gaia}} colour-magnitude diagram, and 119 lie near the WD track. In each case, we investigated the feasibility of a WD being detectable in \textit{GALEX} at their distances. In the case of the 533 spread across the MS and giant tracks, we checked if a log($g$) = 8 CO core WD at 20\,kK would be detectable at the given distances, and found that 509 would be detectable in either the $FUV$ or $NUV$, whilst 30 would not be luminous enough to be detected in either band. We thus flag these 30 systems as being unlikely to be the cause of our observed UV excess (so are thus flagged as `6' in the `Contamination' column of the supplementary table). As for the 509 other systems, it is unclear exactly which source is the origin of the UV excess between our flagged source and this other detection, and these are flagged as `7' in the `Contamination' column of the supplementary table.

For the 119 systems near the WD track, we determined if these systems would be detectable by acquiring what the UV magnitudes of such a WD would be assuming a CO core WD with log($g$) = 8 and the $G$ magnitude of the system. We find that, at their distances, all would be detectable and contribute towards a UV excess. These systems where this other source is the likely cause of our UV excess is flagged as `8' in the `Contamination' column of the supplementary table. Interestingly, in almost all of these cases the contaminating WD has a parallax consistent with our selected UV excess source, indicating that these are likely wide, resolved WD binary systems. We expect contamination from resolved WD binaries accounts for roughly 4 per cent of our sample.

Accounting for all sources of contamination, we estimate that our sample likely has a contamination of between 10-35 per cent, based on how representative the \textsc{Simbad} crossmatch sub-sample is, and how pessimistic we are regarding the 509 systems with another MS system within 5".

\section{Conclusions} \label{sec:Conclusions}

We have compiled a sample of 2597 WD\,+\,FGK binary candidates within 350\,pc from {\textit{Gaia}} by searching for a UV excess within the \textit{GALEX} data by comparing against the MIST models corresponding to the {\textit{Gaia}} parameters for the non-degenerate star. We find that {1642} of these have a low risk of being contaminants. Comparing against \textsc{Simbad} {(Section~\ref{sec:Crossmatching})}, we find 10 systems confirmed by \textsc{Simbad} to be a WD binary (or else a WD binary candidate), and a further 2 WDs or WD candidates not flagged as such within \textsc{Simbad}, but which are found as eclipsing binaries and literature has confirmed as containing a WD or candidate \citep{Lloyd_2007, Zhang_2024, Zhao_2024}. We are able to estimate the ages of some of our systems by either comparing against clusters \citep{Hunt_2024} or by looking for wide WD companions to the system \citep{Shahaf_2024}, though we are able to only do this for a small number of systems within our sample.

In Section~\ref{sec:Method}, we found that {theoretical predictions} significantly under-predict the UV flux of stars across the HRD, which we suspect is a result of the {models} not accounting for the chromosphere. We apply polynomial corrections to these {predictions} in order to bring their magnitudes more in-line with observations.

We estimate a contamination of 10-35 per cent based on our matching against \textsc{Simbad} and other nearby \textit{GALEX} sources (Section~\ref{sec:CompleteContamination}), depending on how representative the \textsc{Simbad} crossmatch is to our full sample. We have expanded our understanding of potential sources of contamination found through our UV excess method, with young stellar objects and wide binaries contributing more significantly as sources of contamination that expected. We cannot gather a full idea of the completeness of our sample without a better understanding of how these systems form an evolve, however we know that our sample is bias towards lower mass, hotter more luminous WDs {(not necessarily reflected in our Case (ii) and Case (iii) systems, see section~\ref{sec:WDParams})}, and we determine the maximum temperature MS star for which our method is effective for detecting WDs next to for given WD parameters, finding that we struggle to detect any WDs in the $FUV$ below a $T_\mathrm{eff, WD}$ of 10,000\,K regardless of the $T_\mathrm{eff, MS}$ of the companion. Whilst we can detect WDs at lower $T_\mathrm{eff, WD}$s (down to $\approx$\,8,000\,K), we find that the hotter WDs are less visible in the $NUV$ than in the $FUV$, with a lower $T_\mathrm{eff, MS}$ being able to mask the WD in the $NUV$.

This sample should prove useful in finding populations of WD\,+\,FGK binaries in the local area to study binary evolution, though follow-up spectroscopic observations are required in order to confirm the presence of a WD. 

\section*{Acknowledgements}

The Digitized Sky Surveys were produced at the Space Telescope Science Institute under U.S. Government grant NAG W-2166. The images of these surveys are based on photographic data obtained using the Oschin Schmidt Telescope on Palomar Mountain and the UK Schmidt Telescope. The plates were processed into the present compressed digital form with the permission of these institutions.

The National Geographic Society - Palomar Observatory Sky Atlas (POSS-I) was made by the California Institute of Technology with grants from the National Geographic Society.

The Second Palomar Observatory Sky Survey (POSS-II) was made by the California Institute of Technology with funds from the National Science Foundation, the National Geographic Society, the Sloan Foundation, the Samuel Oschin Foundation, and the Eastman Kodak Corporation.

The Oschin Schmidt Telescope is operated by the California Institute of Technology and Palomar Observatory.

The UK Schmidt Telescope was operated by the Royal Observatory Edinburgh, with funding from the UK Science and Engineering Research Council (later the UK Particle Physics and Astronomy Research Council), until 1988 June, and thereafter by the Anglo-Australian Observatory. The blue plates of the southern Sky Atlas and its Equatorial Extension (together known as the SERC-J), as well as the Equatorial Red (ER), and the Second Epoch [red] Survey (SES) were all taken with the UK Schmidt.

All data are subject to the copyright given in the copyright summary \url{https://archive.stsci.edu/dss/copyright.html}. Copyright information specific to individual plates is provided in the downloaded FITS headers.

Supplemental funding for sky-survey work at the ST ScI is provided by the European Southern Observatory. 

SGP acknowledges the support by the Science and Technology Facilities Council (grant ST/B001174/1).

Co-funded by the European Union (ERC, CompactBINARIES, 101078773). Views and opinions expressed are however those of the author(s) only and do not necessarily reflect those of the European Union or the European Research Council. Neither the European Union nor the granting authority can be held responsible for them.

FLV acknowledges support from ANID FONDECYT Postdoctoral Grant No. 3250464.

RR acknowledges support from Grant RYC2021-030837-I, funded by MCIN/AEI/ 10.13039/501100011033 and by “European Union NextGeneration EU/PRTR”. This research was partially supported by the AGAUR/Generalitat de Catalunya grant SGR-386/2021 and the Spanish MINECO grant, PID2023-148661NB-I00. 

Juanjuan Ren thanks the support of the China Manned Space Program (Grant No. CMS-CSST-2025-A19)

\section*{Data Availability}

{\textit{Gaia}} data release 3 is available publicly online at \url{https://www.cosmos.esa.int/web/gaia/dr3}. Our sample of excess objects will be made available upon publication on CDS. The full {\textit{Gaia}}-\textit{GALEX} cross match will be made available from the corresponding author on reasonable request.
 



\bibliographystyle{mnras}
\bibliography{GaiaGalex} 




\appendix

\section{Supplementary Figures}
\label{sec:SupFigs}

\begin{figure*}
\centering
	\includegraphics[width = \linewidth]{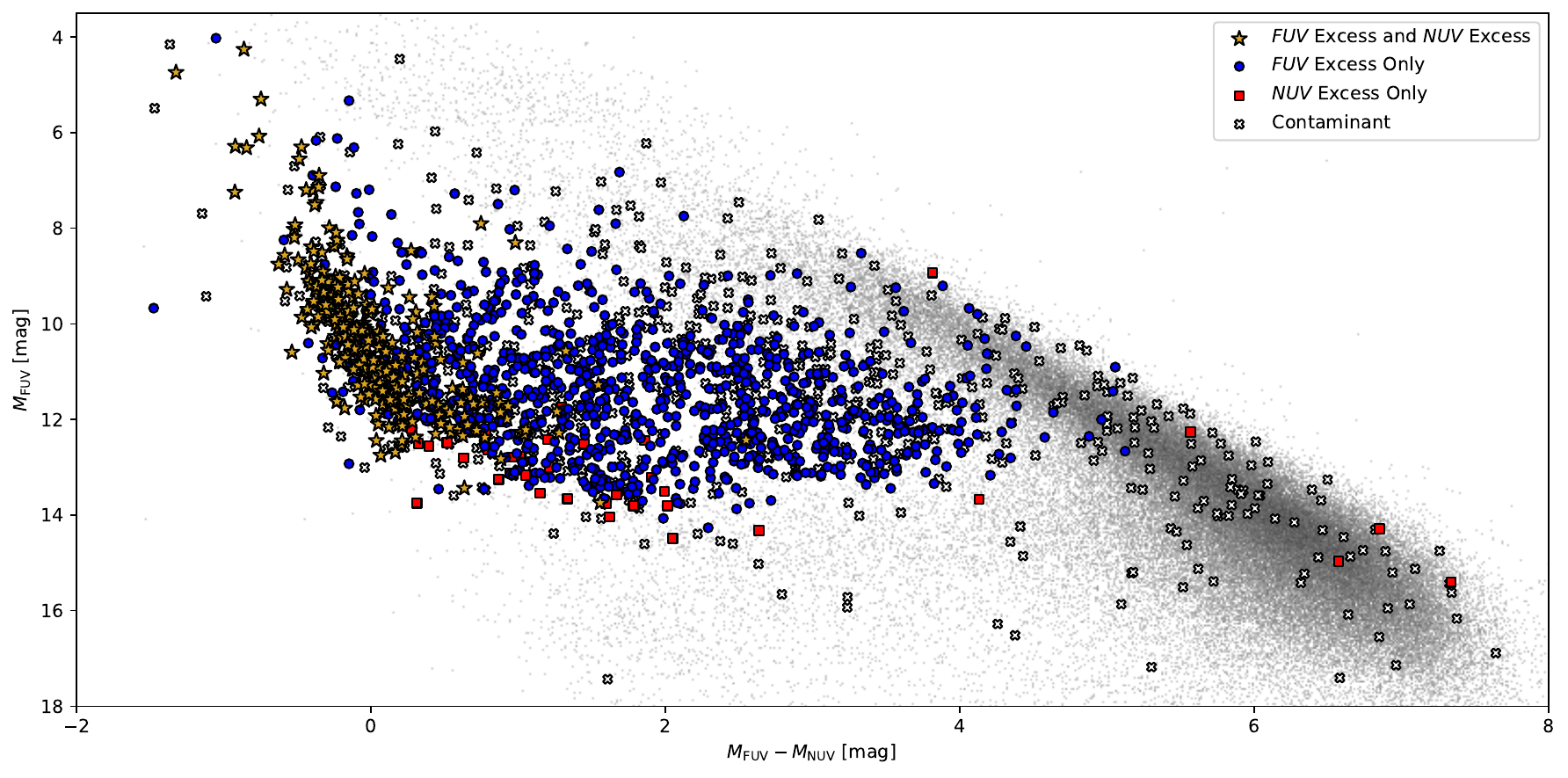}
	\caption{UV colour magnitude diagram displaying where our excess systems lie in and UV  colour space. Our systems with an excess in both UV bands are shown as golden stars, where there was only a $FUV$ excess as blue circles, systems with only an $NUV$ excess as red squares and likely contaminants as a white cross. These `Contaminants' are flagged based on the `Contaminants' column in the supplementary table.}
	\label{fig:ExcessUVCMD}
\end{figure*}

\begin{figure*}
\centering
    \includegraphics[width =\linewidth]{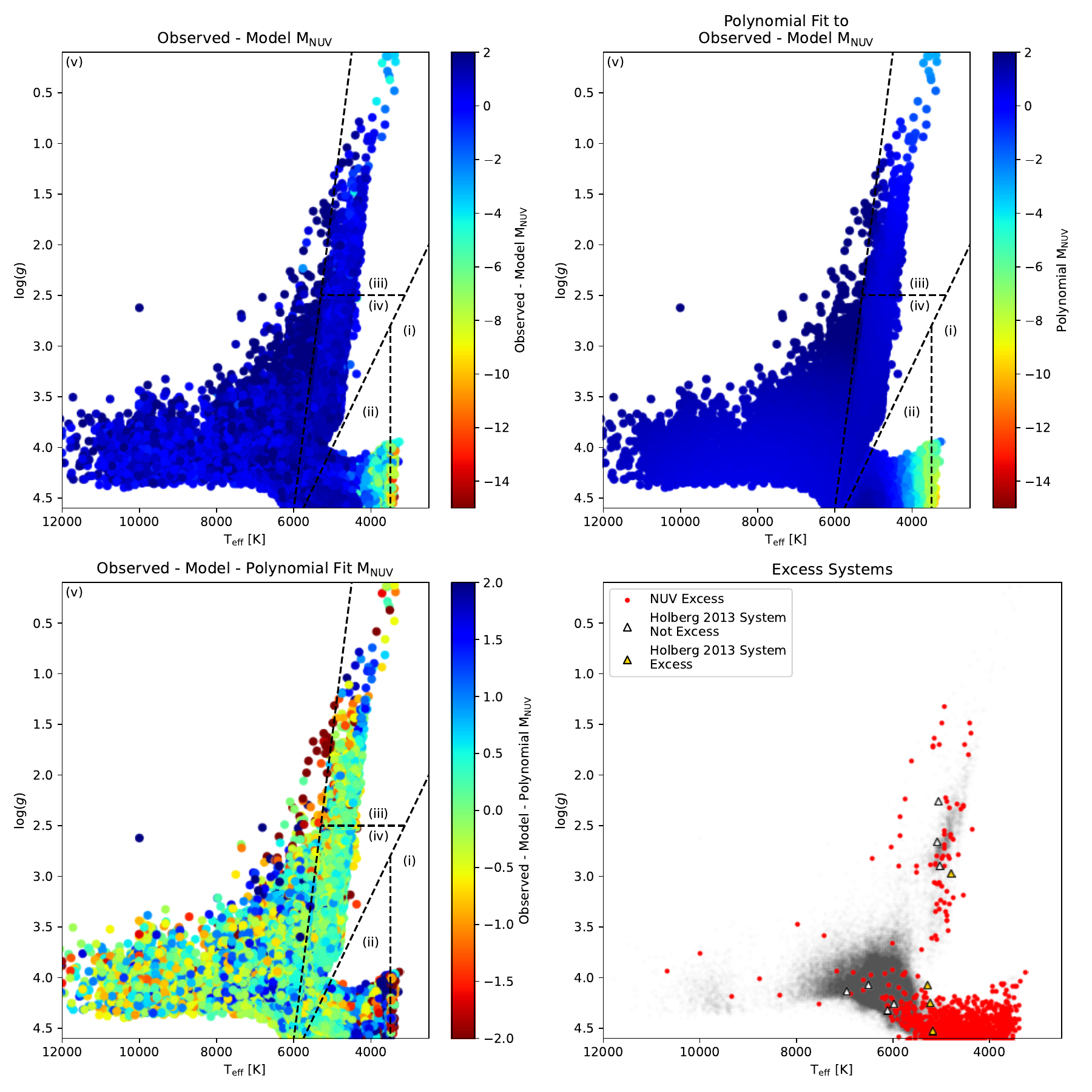}
    \caption{Kiel diagrams of the 488,148 objects in {\textit{Gaia}} DR3 with measured parameters, within 350\,pc and with a $NUV$ detection in \textit{GALEX}. The top left panel is colour-coded by the observed, absolute de-reddened $NUV$ magnitude minus the MIST model magnitude of the system as found by fitting the $T_\mathrm{eff}$, log($g$) and $\mathrm{Fe}/H$ of the system. The top right panel is colour-coded to show the polynomial fit to the distribution, and the bottom left shows the residuals after the polynomial corrections are subtracted. The dashed lines indicate the boundaries of the different regions where different polynomials were fit. The bottom right panel shows our systems that were flagged as an excess, with the known WD\,+\,FGK systems from \citet{Holberg_2013} shown as triangles, the gold triangles where these were found as an excess in this band, and the white triangle where they were not flagged as an excess in this band.}
    \label{fig:loggTeffSpaceNUV}
\end{figure*}

\section{Method Means}
\label{sec:MethodMeans}

\begin{table*}
	\centering
	\caption{{The means and standard deviations of different WD properties based on their fitting methods. Owing to their inaccuracies on individual scales, the individual properties from cases (iv) and (v) are not included in the supplementary table, though their spread is mentioned here.}}
	\begin{tabular}{@{}llll@{}}
	\hline
	Fitting Method & log($g$)$_\mathrm{WD}$ & $T_\mathrm{eff, WD}$ & $M_\mathrm{WD}$\\
	\hline
	(i) Both Fit & 7.63$\pm$0.37 & 20100$\pm$9700 & 0.52$\pm$0.16\\
	(ii) Both Data, $FUV$ Excess Only & 8.04$\pm$0.08 & 17000$\pm$6600 & 0.71$\pm$0.12\\
	(iii) Both Data, $NUV$ Excess Only & 8.06$\pm$0.10 & 16300$\pm$4100 & 0.87$\pm$0.21\\
	(iv) $FUV$ Data Only & 8.00$\pm$0.02 & 16100$\pm$5400 & 0.68$\pm$0.02\\
	(v) $NUV$ Data Only & 7.97$\pm$0.08 & 25900$\pm$12800 & 0.63$\pm$0.11\\
	\end{tabular}
	\label{tab:MethodMeans}
\end{table*}

\section{Crossmatch Tables}
\label{sec:SupTables}

\begin{table*}
	\centering
	\caption{Our candidate systems that are flagged as being potential cluster members in \citet{Hunt_2024}. We do not make cuts based on their probability of membership, as binary evolution can make this less straightforward (e.g. V* V471 Tau being a confirmed member of the Hyades cluster, but only having a probability of 52 per cent)}
	\begin{tabular}{@{}llll@{}}
	\hline
	\textsc{Simbad} Main ID & Cluster & Cluster Age (Myr) & Probability (per cent)\\
	\hline
    2MASS J22131219+7332585 & Collinder 471 & 4.8 & 100 \\
    CHX 18N & Chamaleon I & 5.7 & 100 \\
    Cl* NGC 2632 JC 261 & NGC 2632 & 346.2 & 57 \\
    {\textit{Gaia}} DR2 5579241033805198976 & CWNU 1169 & 79 & 90\\
	Hen 3-545 & Chamaleon I & 5.7 & 98\\
    Hn 23 & Theia 29 & 18.3 & 100 \\
    IRAS 05271+0252 & ASCC 21 & 6.9 & 100 \\
    V* Cl Tau & CWNU 1129 & 15.0 & 100 \\
    V* CV Cha & Chamaleon I & 5.7 & 100 \\
	V* SY Cha & Chamaleon I & 5.7 & 59 \\
	V* V471 Tau & Hyades & 576.8 & 52 \\
	V* VZ Cha & Chamaleon I & 5.7 & 100 \\
	{[K98c]} Em* 109 & Collinder 471 & 4.8 & 100
	\end{tabular}
	\label{tab:ClusterFlag}
\end{table*}	
	
\begin{table*}
    \centering
    \caption{Our \textsc{Simbad} White Dwarf crossmatch results. Here, we give their primary \textsc{Simbad} ID, their likely object type based on the literature, and the source of the classification.}
    \begin{tabular}{@{}lll@{}}
    \hline
    \textsc{Simbad} Main ID & Object Type & Source\\
    \hline

    * 29 Dra & RS CVn (WD + K) & \citet{Hall_1982, McCook_1999}\\

    BD+08 102 & Wide WD\,+\,K (Unlikely CV) & \citet{Barstow_1994, Kellett_1995, Leiner_2018, Sun_2024}\\

    CPD-65 264 & WD\,+\,G & \citet{PathwayI, PathwayV, PathwayVIII}\\

    HD 2133 & WD\,+\,F & \citet{Houk_1975, Burleigh_1997, McCook_1999, Barstow_2001}\\

    HD 18131 & WD\,+\,K & \citet{Vennes_1995A, Vennes_1995B}\\

    HD 90052 & WD\,+\,F & \citet{Burleigh_1997, Holberg_2013}\\
    
    LAMOST J035557.92+432402.3 & WD\,+\,MS & \citet{Ren_2018}\\

    PN A66 35 & WD\,+\,G (Unlikely PN) & \citet{Frew_2008, Tyndall_2013, Jacoby_2021}\\
    
    TYC 110-755-1 & WD\,+\,G & \citet{PathwayV, PathwayVI}\\

    V* V471 Tau & WD\,+\,K & \citet{Nelson_1970}\\
    
    \end{tabular}
    \label{tab:SimbadWDs}
\end{table*}

\section{Sky Images}
\label{sec:SupImages}

\begin{figure*}
    \centering
    \includegraphics[height=0.9\textheight]{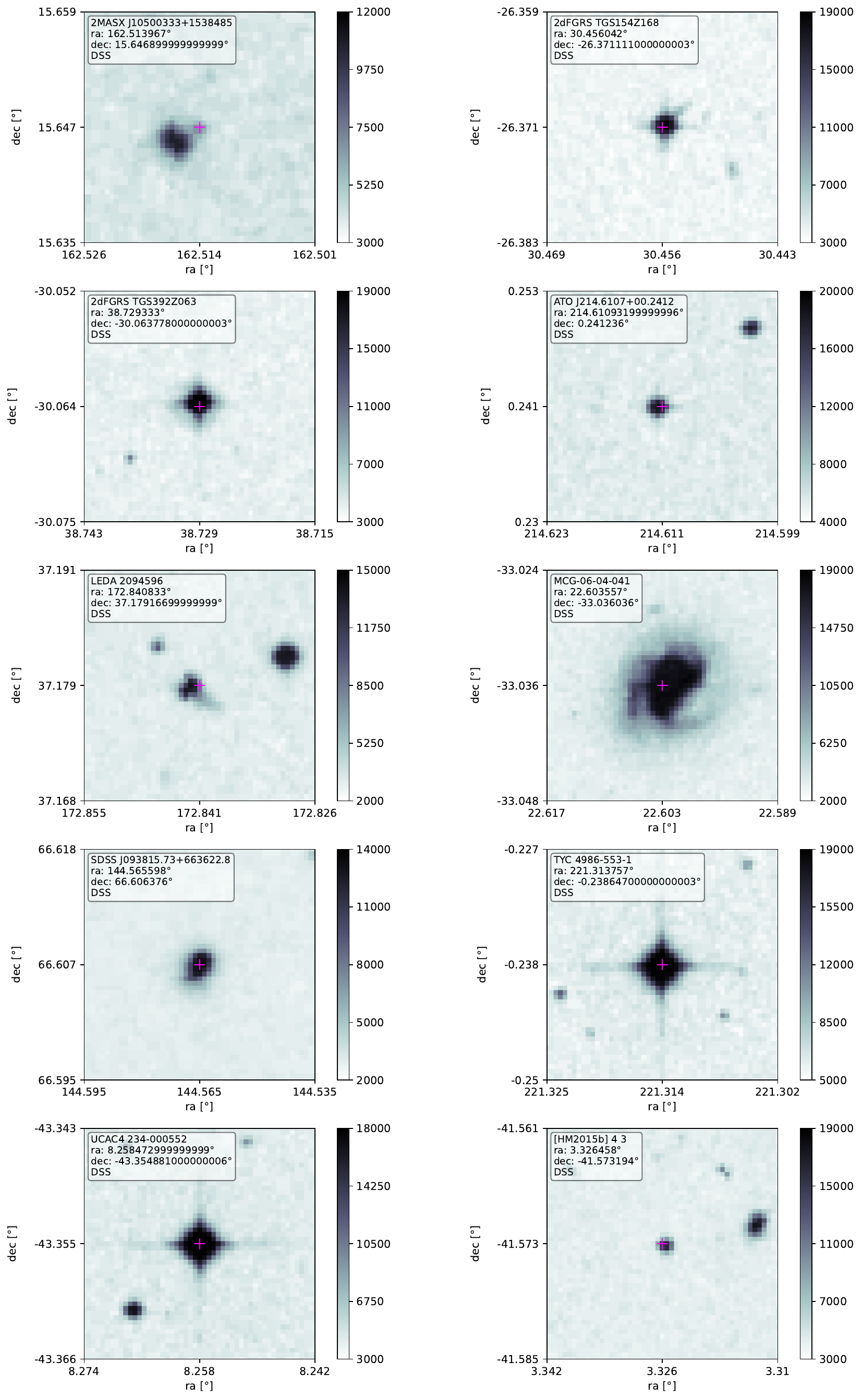}
    \caption{Images of the different systems in our samples that were identified as galaxies in {\sc{Simbad}}. Original images from DSS.}
    \label{fig:Galactic_Images}
\end{figure*}

\begin{figure*}
    \centering
    \includegraphics[width=\linewidth]{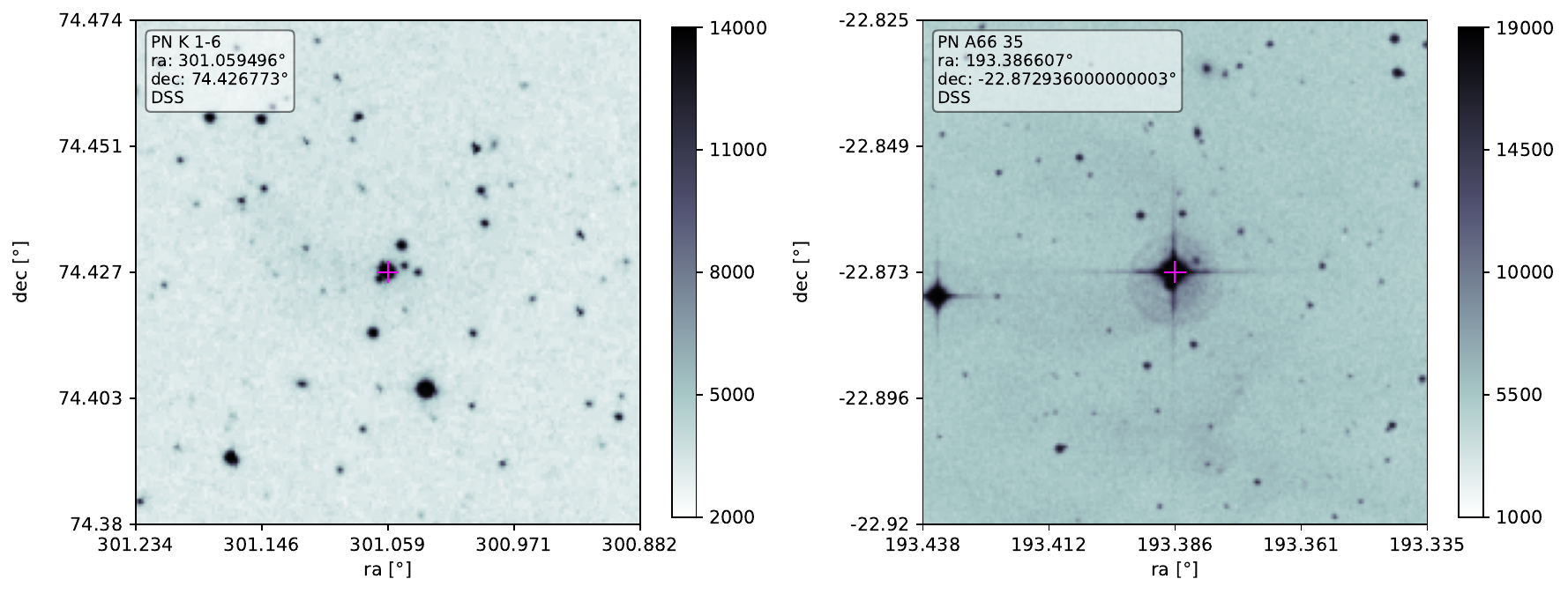}
    \caption{Images of the two systems in our sample flagged as planetary nebulae in \textsc{Simbad}. PN A6635 appears to sit in a cloud of dust here, though as discussed in the text, it is not believed to be a planetary nebulae.}
    \label{fig:PlanetaryNebulae_Images}
\end{figure*}



\bsp	
\label{lastpage}
\end{document}